\documentclass[preprint]{JHEP3} 

\JHEPspecialurl{http://jhep.sissa.it/JOURNAL/JHEP3.tar.gz}
\usepackage{graphicx}
\newcommand\fverb{\setbox\fverbbox=\hbox\bgroup\verb}
\newcommand\fverbdo{\egroup\medskip\noindent%
			\fbox{\unhbox\fverbbox}\ }
\newcommand\fverbit{\egroup\item[\fbox{\unhbox\fverbbox}]}
\newbox\fverbbox

\newcommand{\lb}{{\bar l}}
\newcommand{\qb}{{\bar q}}
\newcommand{\Qb}{{\bar Q}}
\newcommand{\qqgggW}{\qb q ggg W}
\newcommand{\qqQQgW}{\qb q \Qb Q g W}

\newcommand{\e}{\epsilon}
\newcommand{\nn}{\nonumber}
\newcommand{\ba}{\begin{equation}}
\newcommand{\ea}{\end{equation}}
\newcommand{\be}{\begin{eqnarray}}
\newcommand{\ee}{\end{eqnarray}}

\def\ib{{\bar\imath}}

\def\qb{{\bar q}}
\def\tr{\mathop{\rm tr}\nolimits}
\def\si{\sigma}

\def\Gr{\mathop{\rm Gr}\nolimits}

\def\e{\epsilon}

\def\A{{\cal A}}
\def\B{{\cal B}}

\def\tree{{\rm tree}}

\def\oneloop{{1 \mbox{-} \rm loop}}

\title{One-loop amplitudes for ${\bf W+3}$~jet production 
in hadron collisions}

\author{R.~Keith~Ellis and  W.~T.~Giele\\ Fermilab, Batavia, IL 60510, USA }
\author{Zoltan Kunszt \\
Institute for Theoretical Physics, ETH, CH-8093 Z\"urich, Switzerland}
\author{Kirill~Melnikov \\Department of Physics and Astronomy, Johns Hopkins 
University, Baltimore, MD 21218, USA}
\author{Giulia Zanderighi \\Rudolf Peierls Centre for Theoretical Physics, 1 Keble Road, University of
   Oxford, UK}

\received{} 		
\accepted{}		

\preprint{Fermilab-PUB-08-436-T\\
          OUTP-08-17P} 

\abstract{ We employ the recently developed method of generalized
$D$-dimensional unitarity to compute one-loop virtual corrections to
all scattering amplitudes relevant for the production of a $W$ boson
in association with three jets in hadronic collisions, treating all
quarks as massless.  }

\begin{document} 

\section{Introduction}
Physics analyses at the LHC will benefit if accurate predictions for
background and signal processes become available. Arriving at such
predictions often requires next-to-leading order (NLO) QCD
computations. This is particularly true for multi-particle processes
where the tree-level scattering amplitudes involve the strong coupling
constant at a high power. In those cases, changing the renormalization
scale often leads to ${\cal O}(1)$ changes in the cross-section and
more accurate predictions can only be obtained with NLO
computations~\cite{Bern:2008ef}.

The need for NLO corrections to processes with a vector boson and jets
is particularly pressing.  Corrections to vector boson + 1 jet
processes and vector boson + 2 jet processes have been presented in
refs.~\cite{Ellis:1981hk,Arnold:1988dp,Giele:1993dj,Campbell:2002tg}
and have been successfully compared with data in
refs.~\cite{Aaltonen:2007ip,Aaltonen:2007cp}.  The processes
$PP\rightarrow W/Z+N$~jets for $N \geq 3$ have a special
importance. They constitute the principal background to a number of
processes, such as top-pair production and $t$-channel single top
production.  In addition, $PP\rightarrow W/Z+N$~jet production is an
important source of jets + missing energy events, which is often
regarded as a key channel in the search for physics beyond the
Standard Model.

Techniques for NLO computations in the Standard Model in general and
in QCD in particular are well developed. Traditional methods for NLO
calculations are based on the observation that each Feynman diagram
can be represented as a linear combination of tensor integrals.  These
tensor integrals can be reduced to scalar four-, three-, two- and
one-point functions by exploiting Lorentz invariance; this procedure
is known as the Passarino-Veltman reduction technique
\cite{Passarino:1978jh}.

While recent refinements of this 
procedure~\cite{Davydychev:1991va,Duplancic:2003tv,Giele:2004iy,Ellis:2005zh,denner,vanHameren:2005ed,Binoth:2008uq}
have transformed it into a powerful computational tool, there are two
problems inherent in it.  First, the number of diagrams in a
particular process grows faster than $N!$ where  $N$ is the number of
external particles.
While processes with five or more external particles are 
rare at the Tevatron, the increase in energy and luminosity 
of the LHC makes consideration of processes with  $N > 5$~particles 
phenomenologically mandatory.
Second, in the course of the Passarino-Veltman reduction 
procedure for high-point functions, there are numerical instabilities
related to the appearance of Gram determinants. The severity of
this problem also increases with the number of external particles,
because of the concomitant increase in the rank of the integrals.
These two problems make the application of the Passarino-Veltman
reduction technique to processes with more than five external
particles highly
non-trivial~\cite{Denner:2005es,Ellis:2006ss,Bredenstein:2008zb}. For example,
currently there is not a single full process with six external
particles for which NLO QCD corrections are known.

While it may happen that traditional methods of one loop computations
are able to overcome these problems~\cite{Giele:2004ub,denner}, it is
important to develop alternative solutions.  One promising approach is
the method of generalized unitarity that has been developed by Bern,
Dixon, Dunbar and Kosower~\cite{Bern:1994zx}.  Advances by Britto,
Cachazo, Feng~\cite{britto,Britto:2004ap} allowed the development of 
analytic methods for the calculation of the full amplitude, including the
rational part, using recursion
relations~\cite{Bern:2005cq,Berger:2006ci,Berger:2006vq}.  A further
recent advance by Ossola, Pittau and Papadopoulos~\cite{opp} energized
attempts to develop numerical procedures based on
unitarity~\cite{Ellis:2007br,Ossola:2007ax,Berger:2008sj}.  

A new computational
scheme based on $D$-dimensional unitarity has been developed in
Ref.~\cite{Giele:2008ve}. We will refer to this method as generalized
$D$-dimensional unitarity.  
In Refs.~\cite{Giele:2008bc,Ellis:2008ir} the generalized
$D$-dimensional unitarity method was further developed and was shown
to be quite robust. In particular, it was explicitly
demonstrated~\cite{Giele:2008bc} that generalized $D$-dimensional
unitarity is an algorithm of polynomial complexity where the
evaluation time for one-loop pure gluonic amplitudes with $N$ external
particles scales like $N^9$. Moreover, it was also shown that
generalized $D$-dimensional unitarity can be applied to processes with
massive fermions~\cite{Ellis:2008ir}.  The results of these studies
strongly suggest that generalized $D$-dimensional unitarity is an
efficient computational algorithm which is now in a position to have a
phenomenological impact.

The goal of this paper is to make the first steps towards the
application of generalized $D$-dimensional unitarity to phenomenology.
We focus on the computation of virtual one-loop corrections to one of
the important background processes at the Tevatron and LHC for which
the one-loop corrections are still unknown -- the production of the
$W$ boson in association with three jets.  To this end, we have to
consider one-loop corrections to the processes
\be
& 0 \to \bar u +d + g + g + g + W^+,  
\nn \\
& 0 \to \bar u + d +  \bar Q + Q +g + W^+,
\ee
and may assume, without loss of generality, 
 that the quark $Q$ does not  couple to the $W$ boson.
We demonstrate that straightforward application of generalized
$D$-dimensional unitarity allows us to compute {\it all} matrix
elements required for the description of these complicated
processes\footnote{In this article, we do not consider loop
corrections with massive top quarks. Those contributions can be
obtained along the lines described in \cite{Ellis:2008ir}.}.

The paper is organized as follows. In the next Section, we summarize
the salient features of generalized $D$-dimensional unitarity. In
Section~3 we discuss the Dirac algebra and the choices of the
polarization vectors in four- and higher-dimensional space-times. In
Section~4 we describe the computation of all primitive amplitudes
relevant for the process $ 0 \to \bar u d gggW^+$.  In Section~5 we
focus on the amplitudes with four quarks, a gluon and a $W$ boson. We
conclude in Section~6. Numerical results for a specific phase-space
point are collected in Appendix~\ref{app:numres}.

\section{The method}
The method of calculation that we employ in this article is
generalized $D$-dimensional unitarity.  The method relies on the
observation~\cite{Giele:2008ve} that one-loop scattering amplitudes in
QCD can be fully reconstructed once tree-level scattering amplitudes
are known for complex on-shell momenta of external particles,
in, say, six- and eight-dimensions.
The necessity of knowing tree-level
scattering amplitudes in higher-dimensional space-times stems from the
fact that in QCD one-loop amplitudes are divergent and require
regularization. Such regularization is conveniently done by continuing
the dimensionality of space-time from four to $4-2\epsilon$.  At the
end of the calculation, the limit $\epsilon \to 0$ is taken, but
vestiges of the regularization survive as particular finite
contributions (rational terms) in the scattering amplitudes.  It was
pointed out in \cite{Giele:2008ve} that the rational part of the
amplitude can be determined by exploiting the dependence of residues
of one-loop amplitudes on the dimensionality of space-time. Since this
dependence is linear, it is sufficient to know
these residues in two different space-time dimensions to reconstruct
the residue as a function of $D$.

For technical reasons, it is convenient to deal with scattering
amplitudes where external particles are ordered and no permutations
are allowed. Such ordering, for example, automatically fixes the
flavors of all internal lines in the highest-level $N$-point function
that contributes to a particular $N$-particle ordered amplitude.  It
is well-known that such ordering can be achieved without sacrificing
gauge-invariance~\cite{Berends:1987cv,Mangano:1987xk,BG}.  For
tree-level amplitudes, ordering of external particles appears
naturally in color-ordered amplitudes.  For one-loop amplitudes color
ordering does not automatically lead to a complete ordering of all
particles in the amplitude. To achieve this, color-ordered amplitudes
are further decomposed into primitive amplitudes~\cite{Bern:1994fz}.
Those primitive amplitudes can be computed with the help of the
color-stripped Feynman rules~\cite{Bern:1994fz,Bern:1997sc,rules}.
Note, however, that for a given primitive amplitude {\it only}
color-charged particles are ordered while color-neutral particles must
be inserted in all possible locations to achieve a gauge-invariant result. For our
purposes, this implies that the ordering of the $W$ boson is not fixed
and we have to account for all possible insertions of the $W$ bosons
between $\bar u$ and $d$ quarks in a given primitive amplitude.

To define a primitive one-loop amplitude, we employ the following set
of rules:
\begin{itemize}
\item we order all external particles that carry ${\rm SU}(3)$ color charge;

\item we draw a parent diagram with the direction of all fermion lines
fixed such that the loop is always on the right-side of an upwards
oriented fermion line. The order of the external particles is defined
by reading the diagram clockwise. This defines left-handed primitive
amplitudes\footnote{There are also right-handed primitive amplitudes,
where the loop is to the left of an upwards oriented
fermion line. Since left and right primitive amplitudes are
related, in this article we only present left primitive
amplitudes~\cite{Bern:1994fz}.};

\item for an $N$-point scattering process, in general,
the parent must be given by an one-particle irreducible $N$-point
function, represented by a diagram with $N$ propagators in the loop.
For some orderings it may happen that such a parent does not exist, in
this case we draw the diagram by adding dummy lines; 

\item we construct all possible  cuts of a parent and 
we throw away all cuts that contain any dummy line;
\item we process each cut as required by generalized $D$-dimensional
      unitarity;
tree-level on-shell  amplitudes, needed for the 
computation of  residues,  are calculated  
using color-stripped Feynman rules.
\end{itemize}
The parent primitive diagrams that are required for the calculation of
$W+3$~jet amplitudes will be presented later in the paper.

The calculation of residues of primitive amplitudes requires the
knowledge of tree-level amplitudes in six- and eight-dimensional
space-time for complex momenta of external particles. The necessary
matrix elements are constructed by employing Berends-Giele recurrence
relations \cite{BG}. Recall, that these recurrence relations connect
off-shell currents of different multiplicities and with different
particle content. The on-shell scattering amplitudes are obtained from
the on-shell limits of those currents. For the purposes of this paper,
we need to employ currents with up to six on-shell external particles;
a particular example is a fermionic current with three different
fermion flavors and a gluon that contributes to some cuts of the $\bar
u + d + W^+ + Q + \bar Q + g$ scattering amplitude.  We point out
that, in a numerical program, it is possible to define those currents
in a recursive way treating the number of external gluons as a
parameter; currently, this is a necessary, (but not sufficient)
prerequisite for the construction of fully automated computer codes
for NLO QCD computations.

\section{Dirac algebra, spinors and polarization vectors for 
gauge bosons}

\subsection{Four-dimensional case}

For Dirac matrices it is convenient to use the Weyl representation where 
the $\gamma$-matrices are given by 
\begin{equation} \label{eq:gammamatrices}
\gamma^0=\ \left(\matrix{{\bf0}&{\bf1}\cr{\bf1}&{\bf0}\cr}\right)\ ,\,\,
\gamma^i\ =\ \left(\matrix{{\bf0}&-{\bf\sigma}^i\cr
                           {\bf\sigma}^i&{\bf0}\cr}\right)\ ,\,\, 
\gamma^5\ 
\ =\ \left(\matrix{{\bf1}&{\bf0}\cr{\bf0}&-{\bf1}\cr}\right).
\end{equation}

Consider a massless fermion with momentum $p = (E,p_x,p_y,p_z)$ and
let $p_+ = E+p_z$. Solving the Dirac equation for massless quarks, 
we find the following solutions 
\be
u_{\lambda = 1}(p) =
\left (
\begin{array}{c}
\sqrt{p_+} \\
(p_x + i p_y)/\sqrt{p_+} \\
0 \\
0 \\
\end{array} 
\right ),\;\;\;\;
u_{\lambda = -1}(p) =
\left (
\begin{array}{c}
0\\
0\\
(p_x - i p_y)/\sqrt{p_+} \\
-\sqrt{p_+} \\
\end{array} 
\right ),
\ee
where $\lambda = \pm 1$ refers to fermion helicity. Because $p_+$ 
vanishes for $E = -p_z$, the solution for the fermion 
moving in the $-z$ direction requires care. Taking the limit, 
we arrive at 
\be
u_{\lambda = 1}(p) =
\left (
\begin{array}{c}
0 \\
\sqrt{2 E}\\
0 \\
0 \\
\end{array} 
\right ),\;\;\;\;
u_{\lambda = -1}(p) =
\left (
\begin{array}{c}
0\\
0\\
\sqrt{2E} \\
0 \\
\end{array} 
\right ).
\ee
It is easy to see that in the massless case, the anti-particle
solutions of the Dirac equation are related to the particle solutions
so that $v_{\lambda}(p) = u_{-\lambda}(p)$.

The polarization vectors for massless gauge bosons in four dimensions
are also well known.  We present them here for completeness. For a
gluon with momentum
$$
p = E(1,\sin \theta \cos \phi, \sin \theta \sin \phi, \cos \theta),
$$
the polarization vector reads 
\be
\epsilon_{\lambda}(p) = \frac{1}{\sqrt{2}} 
\left ( 0 , \cos \theta \cos \phi - {\rm sgn}(E) \lambda i \sin \phi, 
\cos \theta \sin \phi + {\rm sgn}(E) 
\lambda i \cos \phi, -\sin \theta \right ).
\ee
In this paper we consider outgoing gluons; for this reason, 
all scattering amplitudes are computed with the complex conjugate 
vector $\epsilon^*_\lambda(p)$.

\subsection{$D$-dimensional case}

Generalized $D$-dimensional unitarity requires the knowledge of
tree-level scattering amplitudes in higher-dimensional space-time.  To
compute those amplitudes, we need $D$-dimensional polarization vectors
for gluons, as well as spinors for fermions in $D$ dimensions.
Polarization vectors for gluons were discussed in detail
in~\cite{Giele:2008ve,Giele:2008bc} and we do not repeat that
discussion here.  Weyl fermion spinors in higher-dimensional space-time are
constructed as follows.

We construct a spinor solution for a fermion with light-like momentum
$p$ by using an auxiliary light-like vector $n$ such that $n \cdot p
\neq 0$
\be
u_j (p,n) = \frac{\hat p}{\sqrt{2 p\cdot n}} \chi^{(D)}_j(n),
\;\;\;
\bar u_j (p,n) = \bar \chi^{(D)}_j(n) \frac{\hat p}{\sqrt{2 p\cdot n}}.
\ee
Here $\hat p = p_\mu \Gamma^\mu$, 
where the summed index $\mu$ runs over $D$ components, (the first four of which 
are the $0,x,y,z$) and $\Gamma_\mu$ are the Dirac matrices in $D$ dimensions.
The index $j$ specifies the spinor polarization states.
We choose the $D$-dimensional, $p$-independent spinors
$\chi^{(D)}_j(n)$ in such a way that
\be
\sum_{j=1}^{2^{(D/2-1)}}  \chi^{(D)}_j(n) \otimes \bar \chi^{(D)}_j(n) = \hat n.
\ee
In this case, it is easy to see that the $u_j(p,n)$ spinors 
satisfy both the Dirac equation for massless fermions 
and the completeness relation
\be
\sum_{j=1}^{2^{(D/2-1)}} u_j(p,n) \otimes \bar u_j(p,n) = 
\frac{\hat p \hat n \hat p}{2p\cdot n} = \hat p. 
\ee
We conclude that $u_j(p,n)$ is a valid choice for on-shell
fermion states.

The above construction involves an auxiliary vector $n$ and, for this
reason is quite flexible. Having such a flexibility turns out to be
important, especially since we have to construct on-shell spinors for
complex momenta.  We give a few examples below.

We consider a $D$-dimensional vector $n = (n_0,n_x,n_y,n_z,\{ n_{i \in
(D-4)} \})$, choose $n_0 = 1/2, n_z = 1/2$ and set all other
components to zero.  Then, we need to find the spinors $\chi$ such
that
\be
\sum_{j=1}^{2^{(D/2-1)}}  \chi^{(D)}_j(n) \otimes \bar \chi^{(D)}_j(n) = \hat n
= \frac{1}{2} \left ( \Gamma_0 - \Gamma_z \right ).
\ee
Since $\Gamma_{0,x,y,z}$ are all block-diagonal \cite{collins}, with
``blocks'' being $4 \times 4$ matrices, a $D$-dimensional spinor is
constructed by simple iteration of the four-dimensional
construction. The four-dimensional spinors are given by
\be
\chi^{(4)}_1 = \left (
\begin{array}{c}
1\\
0\\
0\\
0
\end{array}
\right )
,\;\;\;\
\chi^{(4)}_2 = \left ( 
\begin{array}{c}
0\\
0\\
0\\
-1
\end{array}
\right ).
\ee
In six dimensions the eight-component  
spinors are choosen to be
\be
\chi^{(6)}_1 = \left (
\begin{array}{c}
\chi^{(4)}_1\\
0\\
\end{array}
\right )
,\;\;
\chi^{(6)}_2 = \left ( 
\begin{array}{c}
\chi^{(4)}_2\\
0
\end{array}
\right ),
\;\;\;
\chi^{(6)}_3 = \left ( 
\begin{array}{c}
0\\
\chi^{(4)}_1
\end{array}
\right ),
\;\;\;
\chi^{(6)}_4 = \left ( 
\begin{array}{c}
0\\
\chi^{(4)}_2
\end{array}
\right ). 
\label{eq20}
\ee
The case $D=8$ is a simple generalization of the above construction.

We now present two alternative procedures to define fermionic spinors
which we employ when the particular choice of the vector $n$ leads to
numerical instabilities. This occurs for the on-shell momentum $p =
(p_0, 0, 0, p_0)$ since $(p \cdot n)= 0$. To handle this case, we
change the vector $n$ to $n = (1/2,0,0,-1/2,0_{D-4})$ in the above
formulas. However, even this can be insufficient. Indeed, note that
a complex momentum $p = (0,p_x,p_y,0)$ can be light-like. In this
case, we need to choose yet another $n$.  We can take $n =
(1,1,0,0,0_{D-4})$ and choose the following four-dimensional spinors
\be
\chi^{(4)}_1 = \left (
\begin{array}{c}
1\\
1\\
0\\
0
\end{array}
\right )
,\;\;\;
\chi^{(4)}_2 = \left ( 
\begin{array}{c}
0\\
0\\
1\\
-1
\end{array}
\right ).
\ee
The higher-dimensional spinors are obtained from these
four-dimensional solutions along the lines discussed above (see
Eq.(\ref{eq20})).

\section{Processes with two quarks, a $W$ boson and gluons}

In this section we consider the one-loop scattering amplitudes $0 \to
\bar u+d\,+ (n-2)\, g+W^+$.  We refer to $\bar u$ as $\bar q$ and $d$
as $q$ and suppress the label of the $W$ and its decay products in
scattering amplitudes. We note that for a given primitive amplitude
the $W$ boson is inserted in all possible places when the diagram is
traversed in a clockwise direction from $\bar q$ to $q$.

\subsection{Color decomposition of the amplitude}

At tree-level, the $0 \to \bar q + q + (n-2)~{\rm gluons} +W$ scattering 
amplitude can be written as 
\begin{equation}
 {\cal A}_n^\tree(1_{\bar{q}},2_q,3_g,\ldots,n_g)
 \ =\  g^{n-2} \sum_{\sigma\in S_{n-2}}
   (T^{a_{\sigma(3)}}\ldots T^{a_{\sigma(n)}})_{i_2}^{~\ib_1}\
    A_n^\tree(1_{\bar{q}},2_q;{\sigma(3)}_g,\ldots,{\sigma(n)}_g)\ ,
\label{eq:atree}
\end{equation} where $S_{n-2}$ is the permutation group of $(n-2)$
elements and
$A_n^\tree(1_{\bar{q}},2_q;\sigma(3)_g,\ldots,\sigma(n)_g)$ are
color-ordered amplitudes. For all the amplitudes computed in this
paper, we take the $W \bar u d$ interaction vertex to be
$-i\gamma^{\mu} (1-\gamma_5)/2$, so that neither electroweak couplings
nor the Cabibbo-Kobayashi-Maskawa matrix elements are included.  The
$W^+$ decays to $\nu(q_1)+e^+(q_2)$; to account for this, we replace
the polarization vector of the outgoing $W$ by
\ba
\epsilon_{\pm}^\mu=(-1) 
\frac{\bar{u}(q_1) \gamma_\mu \gamma_{\pm} v(q_2)} {(q_1+q_2)^2},\;\; \gamma_{\pm} = \frac{1}{2}(1\pm \gamma_5).
\label{eq:wpol}
\ea
The choice of the polarization vector $\epsilon_{-}$ corresponds to
the $W$ boson interactions in the Standard Model.  The generators of
the ${\rm SU}(3)$ color group are normalized as ${\rm Tr}(T^a T^b) =
\delta^{ab}$ and satisfy the commutation relation
\ba
[T^a, T^b] = -F^c_{ab} T^c\; .
\ea
This normalization 
allows us to employ the color-stripped Feynman rules \cite{Bern:1994fz,Bern:1997sc,rules} 
to calculate color-ordered scattering amplitudes.

\begin{figure}[t!]
\begin{center}
\includegraphics[angle=0,scale=1]{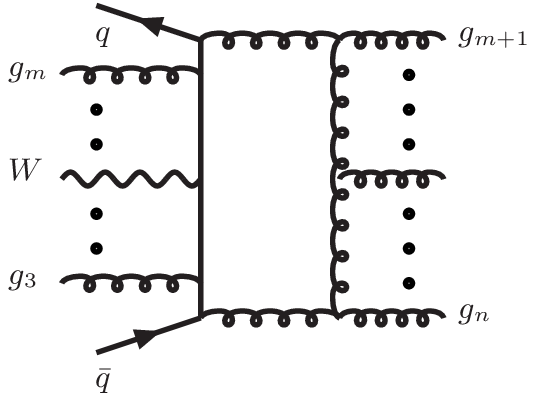}
\includegraphics[angle=0,scale=1]{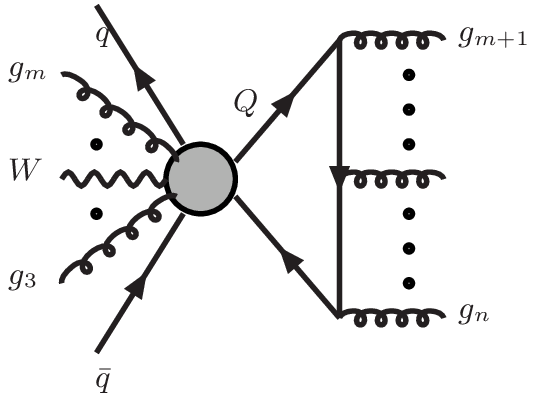}
\caption{Parent diagrams 
for primitive amplitudes $A_n^{L} \left (1_\qb, 3_g,...m_g, 2_q,{m+1}_g
...n_g \right)$ and 
$A_{n}^{L,[1/2]} \left (1_\qb, 3_g,...m_g, 2_q,(m+1)_g
...n_g \right)$.
 All other parent diagrams that contribute to this 
primitive are obtained by considering all possible insertions of the
$W$ boson without changing relative ordering of quarks and gluons.
The shaded circle stands for dummy lines.}
\label{fig1}
\end{center}
\end{figure}

At one-loop, the color decomposition becomes more complicated.
Using the color basis of Ref.~\cite{DelDuca:1999rs}, the 
one-loop scattering amplitude can be written as a linear combination 
of left primitive amplitudes
\begin{eqnarray}
\A_{n}^\oneloop(1_\qb,2_q,3_g,\ldots,n_g)
&=& g^n \biggl[
\sum_{p=2}^n \sum_{\si \in S_{n-2}} 
(T^{x_2} T^{a_{\si_3}} \cdots T^{a_{\si_p}}
 T^{x_1})^{~\bar{i}_1}_{i_2}
(F^{a_{\si_{p+1}}} \cdots F^{a_{\si_n}})_{x_1 x_2} \nn\\
&& \hskip 0.5 cm
\times (-1)^n A_n^{L}(1_\qb,{\si(p)}_g,\ldots,{\si(3)}_g,
2_q,{\si(n)}_g,\ldots,{\si(p+1)}_g)
\label{LoopColorqqNew}\\
&+& {n_f\over N_c} \, \sum_{j=1}^{n-1} \!\sum_{\si\in S_{n-2}/S_{n;j}}
   \!\!\!\Gr^{(\qb q)}_{n;j} (\si_3\ldots,\si_n) 
               A_{n;j}^{[1/2]} (1_{\qb},2_q;{\si(3)}_g,\ldots,{\si(n)}_g) 
\biggr],\nn
\end{eqnarray}
where for $p=2$ the factor ${(T\cdots T)_{i_2}}^{\bar i_1}\rightarrow {(T^{x_2}T^{x_1})_{i_2}}^{\bar i_1}$ 
and for $p=n$ the factor $(F\cdots F)_{x_1x_2}\rightarrow \delta_{x_1x_2}$. 
In the second term $S_{n;j} \equiv {\Bbb  Z}_{j-1}$ 
is the subgroup of $S_{n-2}$ 
that leaves $\Gr^{(\qb q)}_{n;j}$ invariant. 
The color factors read 
\begin{eqnarray}
\Gr^{(\qb q)}_{n;1}(3,\ldots,n) &=& 
N_c (T^{a_3} \cdots T^{a_n})^{~\bar{i}_1}_{i_2} \,, \nn\\
\Gr^{(\qb q)}_{n;2}(3;4,\ldots,n) &=& 0\,, \nn\\
\Gr^{(\qb q)}_{n;j}(3,\ldots,j+1;j+2,\ldots,n) &=&
 \tr(T^{a_3} \cdots T^{a_{j+1}})
  (T^{a_{j+2}} \cdots T^{a_n})^{~\bar{i}_1}_{i_2} \,, 
  \quad j=3,\ldots,n-2, \nn\\
\Gr^{(\qb q)}_{n;n-1}(3,\ldots,n) &=& 
  \tr(T^{a_3} \cdots T^{a_n}) \, \delta^{~\bar{i}_1}_{i_2} \,.
\label{Grqqdef}
\end{eqnarray}
Parent diagrams for primitive amplitudes that involve two quarks and 
gluons are shown in Fig.~\ref{fig1}.

\subsection{Numerical results for $0\to \qqgggW$}

We have extended the {\sf Fortran90} program {\sf Rocket} 
\cite{Giele:2008bc} to include the computation of primitive 
amplitudes with quarks, gluons and gauge vector bosons.  {\sf Rocket}
computes primitive amplitudes in the four-dimensional helicity scheme
\cite{fdh1,fdh2}.  By default, the computation is done with double
precision and, if a particular phase space point is deemed numerically
unstable, it is recomputed with quadruple precision using the package
developed in Ref.~\cite{bailey}. The scalar integrals are evaluated using the 
{\sf QCDLoop} package of ref.~\cite{Ellis:2007qk}.

Note that, since {\sf Fortran}90 supports recursive functions,
implementation of Berends-Giele recursion for an {\it arbitrary}
number of gluons is straightforward and indeed has been done in our
program. Therefore, at least in principle, we can compute one-loop
amplitudes for the process $\bar u d W^+ + n~{\rm gluons}$ where $n$
is an arbitrary number, We have checked that our program produces
gauge-invariant one-loop amplitudes and correct $1/\epsilon^2$ and
$1/\epsilon$ poles for $n$ up to ten. For $n < 3$, we have checked
that our results agree with the known one-loop amplitudes for $W+1$
and $W+2$~jets
\cite{Giele:1993dj,Bern:1994fz,Campbell:1997tv,Bern:1997sc}.  For the
description of $W+3$~jets, we require one-loop amplitudes with five
external partons and this is what we focus on in the remainder of this
paper.

We present numerical results for the primitive amplitudes with seven
external particles at a particular kinematic point considered in
Ref.~\cite{Berger:2008sz}.  The momenta are chosen to be
\begin{eqnarray}
p_{1}(\bar q) 
&=&  {\mu \over 10} (1, \cos\alpha \cos\beta, \cos \alpha \sin \beta,
                                 \sin \alpha )\,, \nn \\
p_{2}(q) &=&  {\mu \over 2}\, (-1, \sin \theta,
           \cos \theta \sin \phi, \cos \theta \cos\phi)\,, \nn\\
p_{3}(g) &=&  {\mu \over 2}\, (-1, - \sin \theta,
         - \cos \theta \sin \phi, - \cos \theta \cos \phi)\,, \nn \\
p_{4}(g) &=&  {\mu\over 3} (1,1,0,0) \,, \nn \\
p_{5}(g) &=&  {\mu\over 8} (1, \cos\beta, \sin \beta,0) \,, \nn \\
p_{6}(e^+) &=& {\mu \over 12} (1, \cos\gamma \cos\beta,
            \cos \gamma \sin \beta, \sin \gamma)\,, \nn \\
p_{7}(\nu) &=& -p_1-p_2-p_3-p_4-p_5-p_6\,,
\label{kinpoint}
\end{eqnarray}
where $\mu = 7$ GeV and 
\begin{eqnarray}
 \theta = {\pi\over 4}\,,\hskip 1 cm
 \phi   =  {\pi\over 6}\,,\hskip 1 cm
 \alpha = {\pi \over 3}\,,\hskip 1 cm
 \gamma = {2 \pi \over 3} \,, \hskip 1 cm
 \cos \beta = - {37\over 128}. 
\end{eqnarray} The momenta $p_{6}$ and $p_7$ are used to define the
polarization vector of the $W$ boson, eq.(\ref{eq:wpol}).

Our results for unrenormalized primitive amplitudes $A_n^{L}$ and
$A_{n}^{L,[1/2]}$ are summarized in the Appendix, see
Tables~\ref{table:nterm3}-\ref{table:nterm6} and
Tables~\ref{table:nf1}-\ref{table:nf2}, respectively. We have checked
that all primitive amplitudes
 have correct divergences and are gauge invariant.
Moreover, we have tested the validity of our results for primitive
amplitudes by reproducing a diagrammatic computation of color-ordered
amplitudes by taking appropriate linear combinations of primitive
amplitudes.  Finally, our program reproduces the results for the
leading-color primitive amplitude $A_5^{L}(1_\qb, 2_q, 3_g,4_g,5_g)$
computed recently in Ref.~\cite{Berger:2008sz}.

\begin{figure}[t]
\begin{center}
\includegraphics[angle=-90,scale=0.35]{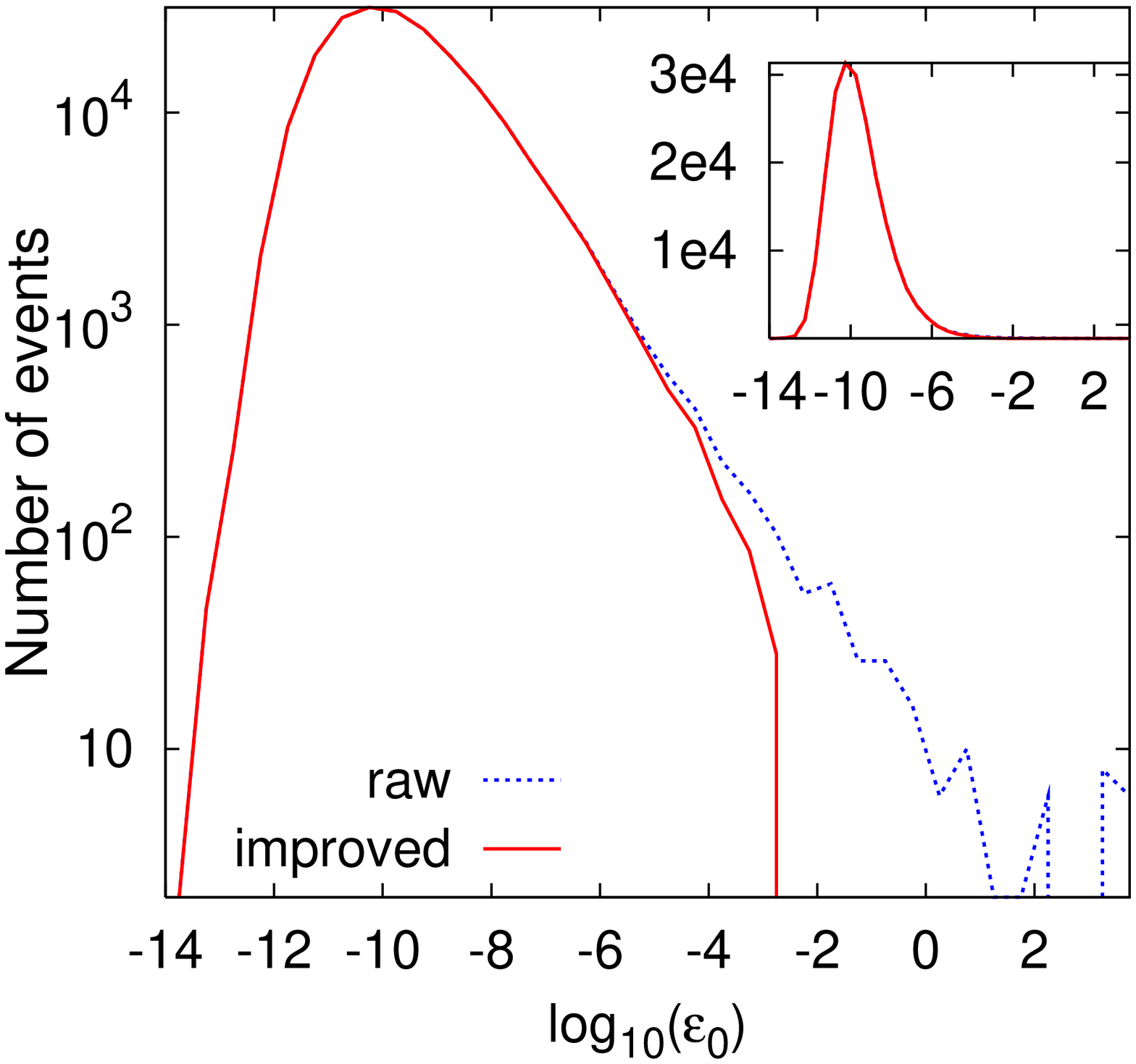}
\includegraphics[angle=-90,scale=0.35]{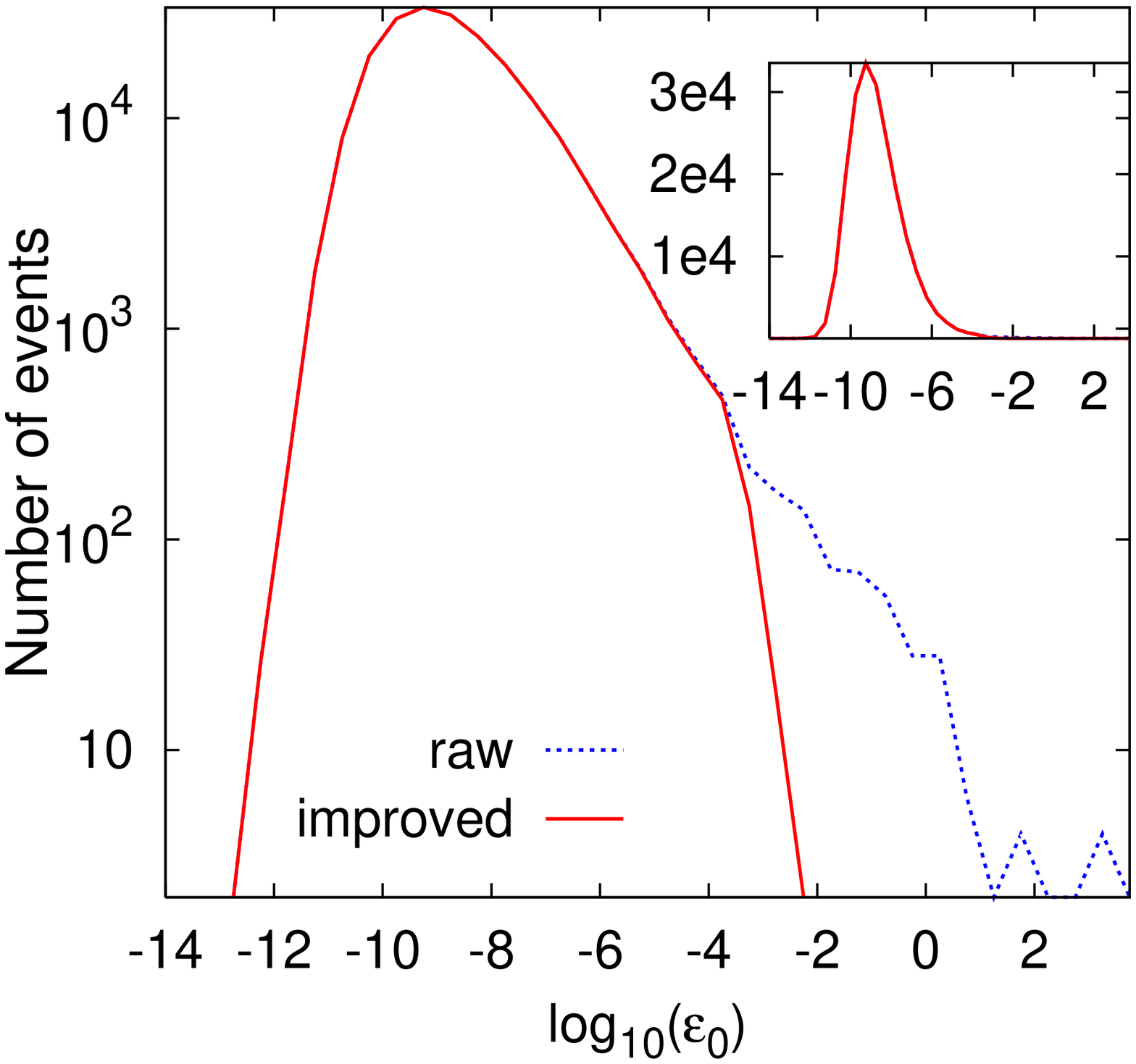}
\caption{Accuracy for 
$A_5^L(1^+_\qb, 2^-_q,3^-_g,4^+_g,5^-_g)$
(left panel)  and for  
$A_5^{L}(1_\qb^+, 3^-_g,4^+_g,5^-_g,2^-_q)$
(right panel)
for $10^5$ randomly generated phase-space points. 
The raw double precision data as well as the result of numerical 
improvements are shown (see text for details). The inset shows the same
plots in a linear scale.}
\label{fig0}
\end{center}
\end{figure}
Next, we address the issue of the numerical stability of the
computation.  As was done earlier for similar studies of gluon
amplitudes, we take care of numerical instabilities by performing
computations with higher precision. Since higher precision slows the
computation, it is desirable to use it only for the phase-space points
that suffer from numerical instabilities.  The question we have to
address therefore is how to detect numerical instabilities.  To study
this, we generate $10^5$ random phase-space points using {\sf
Rambo}~\cite{Kleiss:1985gy} with minimal constraints $ E_\perp >
10^{-2} \sqrt{s}$, $|\eta| < 3$ and 
$\Delta R =\sqrt{\Delta \eta^2+\Delta \phi^2}> 0.4$ 
and calculate the primitive amplitudes for
$0 \to \qqgggW$ with double and quadruple precision.
For each phase-space point, we can check whether or not the double
precision computation of a primitive amplitude reproduces the
analytically known results for double and single poles in $\epsilon$
and if the system of equations is solved with sufficient accuracy for
each residue. To explain the latter test, we remind the reader that
each residue is completely parameterized by a certain number of
coefficients.  We can check how well these coefficients are computed
by choosing a random loop momentum, calculating the residue and
checking how well this residue is obtained from the previously
computed coefficients.  We assign a relative error to each coefficient
following the mismatch in this reconstruction. These errors are used
to estimate the total error in the calculation of the primitive
amplitude. By requiring that the relative precision in the poles and
in the amplitude is better than $10^{-3}$ we find that around 0.3\% of
the points are recomputed in quadruple precision.

After unstable points are recomputed with quadruple precision, the
numerical instabilities are under control.  This is demonstrated in
Fig.~\ref{fig0} for two primitive amplitudes
$A_5^{L}(1^+_\qb,2^-_q,3^-_g,4_g^+,5_g^-)$ and $A_5^{L}(1^+_\qb,
3^-_g,4_g^+,5_g^-,2^-_q)$ where we show the number of events as a
function of the relative accuracy $\epsilon_0$ defined as the absolute
value of the difference between double and quadruple precision
results, divided by the quadruple precision result.
We note, however, that the numerical stability of the amplitudes
illustrated in Fig.~\ref{fig0} is generic, largely independent of the
choice of the primitive amplitude and helicities of quarks and gluons.
In fact, the two amplitudes considered in Fig.~\ref{fig0} are on the
two sides of the spectrum. The leading-color amplitude
$A_5^{L}(1^+_\qb,2^-_q,3^-_g,4_g^+,5_g^-)$ has the {\it minimal}
number of cuts, since the $W$ boson can only be inserted in one place,
between $1_\qb$ and $2_q$. On the contrary, the amplitude
$A_5^{L}(1^+_\qb,3^-_g,4_g^+,5_g^-,2^-_q)$ has the {\it maximal}
number of cuts since the $W$ boson can be inserted in four different
places.  Thus, among all primitive amplitudes with two quarks and
three gluons, maximal computational effort is required for
$A_5^{L}(1^+_\qb,3^-_g,4_g^+,5_g^-,2^-_q)$ so that the issues of numerical
stability may be expected to be worst in this case.

We expect that further optimization of the procedure for identifying
unstable points may be required to arrive at an optimal compromise
between numerical accuracy and speed of the code.
For instance, with an arbitrary precision package such as that
of ref.~\cite{bailey}, one can design a procedure where instead of using
fixed quadruple precision for unstable points, the number of digits in
the higher precision calculation is established according to how
unstable the point is.
We plan to study these issues in the future.

Finally, we remark on the CPU-time required for the evaluation of
one-loop amplitudes. We find that it takes about 45-50~msec to
evaluate the leading-color primitive amplitude
$A_5^L(1_\qb,2_q,3_g,4_g,5_g)$ on a computer with 2.33~GHz Pentium
Xeon processor using the {\sf intel} fortran compiler.  This is
comparable to the evaluation times for six-gluon primitive amplitudes
\cite{Giele:2008bc}. For more complicated primitive amplitudes, the number 
of cuts increases and the evaluation times scale accordingly. For 
example, for $A_5^L(1_\qb,3_g,4_g,5_g,2_q)$, the case with the
maximal number of cuts, 
it takes about 160~msec to compute a single primitive amplitude.

\section{Processes with two quark pairs, a $W$ boson and a gluon}

\subsection{Color decomposition of the amplitude}

\begin{figure}[t!]
\begin{center}
\includegraphics[angle=0,scale=1]{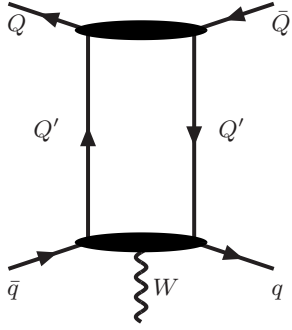}
\caption{Prototype parent diagram for primitive amplitudes 
with four quarks, a $W$ boson and a gluon 
which contain a closed fermion loop.
The gluon can be inserted in four possible ways 
into the prototype graph, leading to four primitive amplitudes. 
Note that the $W$ only couples to $q$. 
The solid blobs denote the dummy lines introduced in section 2.}
\label{fig2}
\end{center}
\end{figure}

\begin{figure}[b]
\begin{center}
\includegraphics[angle=0,scale=1]{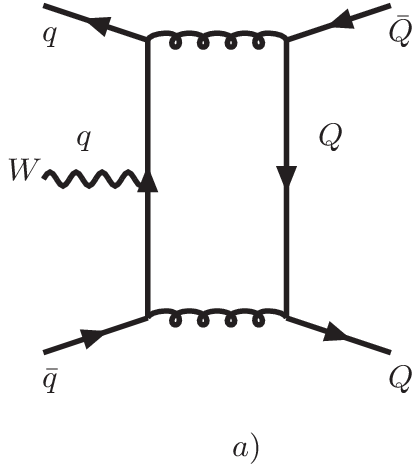}
\includegraphics[angle=0,scale=1]{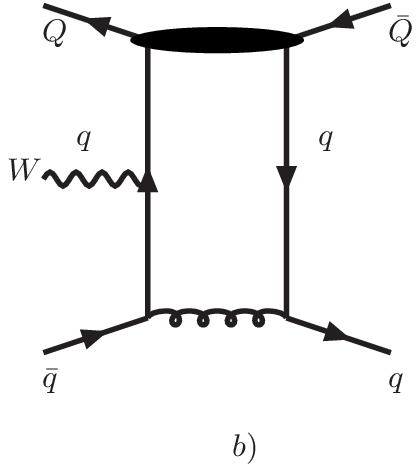}
\includegraphics[angle=0,scale=1]{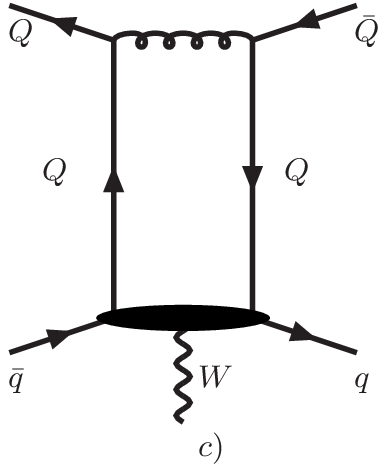}
\caption{Prototype parent 
diagrams for primitive amplitudes with four quarks, a $W$ boson and a
gluon for classes $a,b$ and $c$.  The gluon can be inserted
in four possible ways into any of the prototype graphs, leading to
four primitive amplitudes in each case. For the diagrams in class
$b$, all possible insertions of a $W$ boson to a given parent
primitive should be considered; note that the $W$ only couples to
$q$. The solid blobs denote the dummy lines introduced in section 2.}
\label{fig3}
\end{center}
\end{figure}

We now turn to processes with two quark pairs, the $W$ boson and a
gluon, $0 \to \bar u + d + \Qb + Q + W + g$. We will again refer to
$\bar u$ as $ \bar q$ and to $d$ as $q$.  To construct color-ordered
primitive amplitudes, we assume that the $W$ boson can not couple
to the quark $Q$.  The color decomposition of the four-quark and
one-gluon amplitude at tree-level reads
\begin{eqnarray}
\B^\tree (1_\qb,2_q,3_\Qb,4_Q,5_g) &=&  g^3 \Bigg[
(T^{a_5})_{i_4}^{~\ib_1} \delta_{i_2}^{~\ib_3} B^\tree_{7;1}
 +\frac{1}{N_c}
(T^{a_5})_{i_2}^{~\ib_1} \delta_{i_4}^{~\ib_3} 
B^\tree _{7;2} \nn \\
&+& (T^{a_5})_{i_2}^{~\ib_3} \delta_{i_4}^{~\ib_1} 
B^\tree _{7;3}
    +\frac{1}{N_c} (T^{a_5})_{i_4}^{~\ib_3} 
\delta_{i_2}^{~\ib_1}B^\tree _{7;4}\Bigg],
\end{eqnarray}
in an obvious notation where $a_5$ is the color index of the produced gluon.
The color decomposition of the four-quark and one-gluon amplitude at
one loop reads
\begin{eqnarray}
\B^\oneloop (1_\qb,2_q,3_\Qb,4_Q,5_g) &=&  g^5 \Bigg[
N_c (T^{a_5})_{i_4}^{~\ib_1} \delta_{i_2}^{~\ib_3} B_{7;1}
 +(T^{a_5})_{i_2}^{~\ib_1} \delta_{i_4}^{~\ib_3} 
B_{7;2} \nn \\
&+&N_c (T^{a_5})_{i_2}^{~\ib_3} \delta_{i_4}^{~\ib_1} 
B_{7;3}
    +(T^{a_5})_{i_4}^{~\ib_3} 
\delta_{i_2}^{~\ib_1}B_{7;4}\Bigg].
\end{eqnarray}
Each of these one-loop color-ordered amplitudes can be further written as 
\be
B_{7;i} = B^{[1]}_{7;i}+ \frac{n_f}{N_c} B^{[1/2]}_{7;i},\qquad i=1,2,3,4\,, 
\ee
to separate the diagrams with a closed fermion loop from the other ones.
The amplitudes $B^{[1]}_{7;i}$ and $B^{[1/2]}_{7;i}$ can be written as
linear combinations of primitive amplitudes.  Those primitive
amplitudes are shown in Figs.~\ref{fig2} and \ref{fig3}.

For the amplitudes with an additional closed 
fermion loop we find 
\begin{eqnarray}
&& B^{[1/2]}_{7;1} = -A^{[1/2]}_L(1_\qb, 5_g,4_Q,3_\Qb,2_q), \nn \\
&& B^{[1/2]}_{7;2} = -A^{[1/2]}_L(1_\qb, 4_Q,3_\Qb,2_q,5_g), 
\nn \\
&& B^{[1/2]}_{7;3} = -A^{[1/2]}_L(1_\qb, 4_Q,3_\Qb,5_g,2_q), 
\nn \\
&& B^{[1/2]}_{7;4} = -A^{[1/2]}_L(1_\qb, 4_Q,5_g,3_\Qb,2_q). 
\end{eqnarray}

The three classes of primitive amplitudes that we need to consider 
for four-quark processes without closed fermion loops are shown in Fig.~\ref{fig3},
\be
B^{[1]}_{7;i} = B^{[1],a}_{7;i}+
B^{[1],b}_{7;i} + B^{[1],c}_{7;i}\,.
\ee
Amplitudes from each class are written as linear combinations of
primitives amplitudes.

For class $a$, we find 
\begin{eqnarray}
B^{[1],a}_{7;1} &=& 
\left ( 1- \frac{1}{N_c^2} \right ) 
A_L^{[1],a}\left (1_\qb, 2_q, 3_\Qb, 4_Q, 5_g \right )
- \frac{1}{N_c^2}
\left (
-A_L^{[1],a}\left (1_\qb, 5_g, 2_q, 3_\Qb, 4_Q \right )
\right.  \\
&& \left .
-A_L^{[1],a} \left (1_\qb, 5_g,2_q,  4_Q, 3_\Qb  \right )
-A_L^{[1],a} \left (1_\qb,  2_q,5_g, 3_\Qb, 4_Q \right )
-A_L^{[1],a} \left (1_\qb, 2_q, 5_g, 4_Q,3_\Qb  \right )
\right. \nn \\
&& \left .
-A_L^{[1],a} \left (1_\qb, 2_q, 3_\Qb,5_g,  4_Q  \right )
-A_L^{[1],a} \left (1_\qb, 2_q,  4_Q, 5_g, 3_\Qb \right )
+A_L^{[1],a} \left (1_\qb, 2_q, 4_Q, 3_\Qb , 5_g \right )
\right ), \nn \\
B^{[1],a}_{7;2} &=& 
+A_L^{[1],a} \left (1_\qb, 2_q, 5_g, 4_Q,3_\Qb  \right )
- A_L^{[1],a} \left (1_\qb, 2_q,  4_Q,5_g,3_\Qb  \right )
+ A_L^{[1],a} \left (1_\qb, 2_q,  4_Q,3_\Qb,5_g  \right )
\nn \\
&& 
-\frac{1}{N_c^2} \left ( 
A_L^{[1],a} \left (1_\qb, 5_g, 2_q, 3_\Qb,4_Q  \right )
+ A_L^{[1],a} \left (1_\qb, 5_g, 2_q, 4_Q, 3_\Qb  \right )
\right ),
\\
B^{[1],a}_{7;3} &=& 
 \left ( 1- \frac{1}{N_c^2} \right ) 
A_L^{[1],a}\left (1_\qb, 2_q, 5_g, 3_\Qb, 4_Q \right )
- \frac{1}{N_c^2}
\left (
-A_L^{[1],a}\left (1_\qb, 5_g, 2_q, 3_\Qb, 4_Q \right )
\right.  \\
&& \left .
-A_L^{[1],a} \left (1_\qb, 5_g,2_q,  4_Q, 3_\Qb  \right )
-A_L^{[1],a} \left (1_\qb,  2_q, 3_\Qb, 5_g, 4_Q \right )
-A_L^{[1],a} \left (1_\qb, 2_q, 4_Q,5_g, 3_\Qb  \right )
\right. \nn \\
&& \left .
-A_L^{[1],a} \left (1_\qb, 2_q, 3_\Qb,  4_Q,5_g  \right )
-A_L^{[1],a} \left (1_\qb, 2_q,  4_Q, 3_\Qb,5_g \right )
+A_L^{[1],a} \left (1_\qb, 2_q,5_g, 4_Q, 3_\Qb\right )
\right ), \nn \\
B^{[1],a}_{7;4} &=& 
- A_L^{[1],a} \left (1_\qb, 5_g, 2_q, 4_Q,3_\Qb  \right )
- A_L^{[1],a} \left (1_\qb, 2_q, 5_g,  4_Q,3_\Qb  \right )
- A_L^{[1],a} \left (1_\qb, 2_q,  4_Q,3_\Qb,5_g  \right )
\nn \\
&& 
-\frac{1}{N_c^2} \left ( 
A_L^{[1],a} \left (1_\qb,  2_q,4_Q, 5_g,
 3_\Qb  \right )
+ A_L^{[1],a} \left (1_\qb, 2_q,3_\Qb,5_g, 4_Q  \right )
\right ).
\end{eqnarray}
Note that in this formula there are  amplitudes where fermions and 
anti-fermions alternate $(\qb q \Qb Q)$ 
and amplitudes where this is not the case $(\qb q Q \Qb)$.
The latter can be reduced to the former using the following 
$C$-parity relation 
\begin{eqnarray}
A_L^{[1],a} \left (1_\qb,\ldots, 2_q,\ldots,  4^{\lambda_4}_{Q},\ldots, 3^{\lambda_3}_\Qb,\ldots\right )
= (-1)^{n+1}  
A_L^{[1],a} \left (1_\qb, \ldots,2_q,\ldots  4^{\lambda_4}_\Qb,\ldots, 
3^{\lambda_3}_{Q},\ldots  \right ). \nonumber \\
\end{eqnarray}
Here $n$ is the number of {\it external} gluons sandwiched 
between the $\bar Q$  and $Q$ spinors.

For classes $b$ and $c$ we obtain 
\begin{eqnarray}
B^{[1],b}_{7;1} &=& 
\frac{1}{N_c^2} 
A_L^{[1],b} \left (1_\qb, 5_g, 4_Q,3_\Qb, 2_q  \right ),
\nn \\
B^{[1],b}_{7;2} &=& 
-\frac{1}{N_c^2} 
\left ( 
A_L^{[1],b} \left (1_\qb, 5_g, 4_Q,3_\Qb, 2_q  \right )
+A_L^{[1],b} \left (1_\qb, 4_Q,5_g,3_\Qb, 2_q  \right )
+A_L^{[1],b} \left (1_\qb, 4_Q, 3_\Qb,5_g,2_q  \right )
\right ),
\nn \\
B^{[1],b}_{7;3} &=& 
\frac{1}{N_c^2} 
A_L^{[1],b} \left (1_\qb, 4_Q,3_\Qb,5_g, 2_q  \right ),
\nn \\
B^{[1],b}_{7;4} &=& 
-A_L^{[1],b} \left (1_\qb, 5_g, 4_Q,3_\Qb, 2_q  \right )
-A_L^{[1],b} \left (1_\qb, 4_Q,3_\Qb,5_g, 2_q  \right )
-A_L^{[1],b} \left (1_\qb, 4_Q,3_\Qb, 2_q,5_g  \right )
\nn \\
&& 
-\left ( 1 - \frac{1}{N_c^2}
\right )
A_L^{[1],b} \left (1_\qb, 4_Q,5_g,3_\Qb, 2_q  \right ).
\end{eqnarray}

\begin{eqnarray}
B^{[1],c}_{7;1} &=& 
\frac{1}{N_c^2} 
A_L^{[1],c} \left (1_\qb, 5_g, 4_Q,3_\Qb, 2_q  \right ),
\nn \\
B^{[1],c}_{7;2} &=& 
-A_L^{[1],c} \left (1_\qb,5_g, 4_Q, 3_\Qb,2_q \right )
-A_L^{[1],c} \left (1_\qb, 4_Q,5_g,3_\Qb, 2_q  \right )
-A_L^{[1],c} \left (1_\qb, 4_Q,3_\Qb, 5_g,2_q  \right )
\nn \\
&& 
-\left (1-\frac{1}{N_c^2} \right )
A_L^{[1],c} \left (1_\qb, 4_Q,3_\Qb, 2_q,5_g  \right ),
\nn \\
B^{[1],c}_{7;3} &=& 
\frac{1}{N_c^2} 
A_L^{[1],c} \left (1_\qb, 4_Q,3_\Qb,5_g, 2_q  \right ),
\\
B^{[1],c}_{7;4} &=& 
-\frac{1}{N_c^2} \left ( 
A_L^{[1],c} \left (1_\qb, 5_g, 4_Q,3_\Qb, 2_q  \right )
+A_L^{[1],c} \left (1_\qb, 4_Q,3_\Qb,5_g, 2_q  \right )
+A_L^{[1],c} \left (1_\qb, 4_Q,3_\Qb, 2_q,5_g  \right )
\right ), \nn
\end{eqnarray}

\begin{figure}[t]
\begin{center}
\includegraphics[angle=-90,scale=0.45]{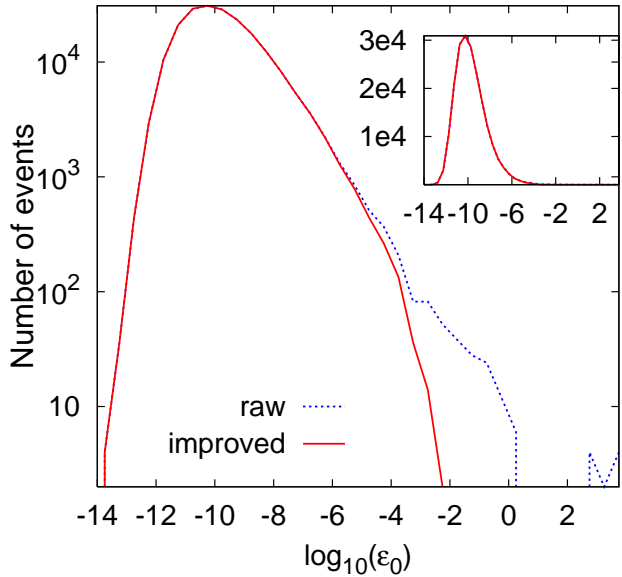}
\caption{Accuracy for $A_L^{[1],a}(1^+_\qb, 2^-_q,5_g^+, 3^-_Q,4^+_\Qb)$ amplitude for
$10^5$ randomly generated phase-space points.  The raw double
precision data as well as the result of numerical improvements are
shown. The inset shows the same plots in a linear scale.}
\label{fig:ferm}
\end{center}
\end{figure}

\subsection{Numerical results for $0 \to \qqQQgW$ amplitudes} 

We now present numerical results for $0 \to \qqQQgW$
amplitudes. We use the kinematic point in Eq.~(\ref{kinpoint}),
where the momentum $p_3$ is the momentum of $\bar Q$ and the momentum
$p_4$ refers to $Q$. The results of the calculation are given in
Tables~\ref{tab7}-\ref{tab10}. The results for primitive amplitudes given
in these Tables are sufficient to obtain numerical results for the
color-ordered amplitudes $B_{7;i},\;i=1,\dots 4$ using the equations
given in the previous subsection. In fact, we have checked our results
for the primitive amplitudes by calculating the color-ordered
amplitudes $B_{7;i}$, $i=1, \ldots, 4$ diagrammatically and verifying
that linear combinations of the primitive amplitudes computed with
{\sf Rocket} reproduce results of the diagrammatic computation.  The
numerical stability of a four-quark amplitude is illustrated in
Fig.~\ref{fig:ferm}.  It is very similar to the numerical stability of
the amplitudes with two quarks and three gluons discussed earlier in
detail and to other four-quark amplitudes.

\section{Conclusions} 

In this paper we have shown that a straightforward application of the
generalized $D$-dimensional unitarity method proposed in
ref.~\cite{Giele:2008ve} allows us to compute all one-loop scattering
amplitudes required to describe the production of a $W$ boson in
association with three jets at hadron colliders.  We observe
satisfactory performance in terms of numerical stability and required
run times. We feel confident that the results of this paper provide a
solid foundation for computing the one-loop virtual corrections to the
production of the $W+3$~jets in hadron collisions. 

On a more general side, the current version of {\sf Rocket} computes 
one-loop amplitudes for processes 
$0 \to ~n~{\rm gluons}$, $0 \to \qb q +n~{\rm gluons}$, 
$0 \to \qb q W +n ~{\rm gluons}$ and $0 \to \qb  q \Qb Q W + 1~{\rm gluon}$.
It is straightforward to extend the program to include similar processes 
with the $Z$ boson and processes with massive quarks 
$0 \to \bar{t}t  +n~{\rm gluons}$. This list is a testimony to the power of the method 
and indicates that the development of automated programs for one-loop calculations 
may finally be within reach.

\vskip 0.4cm
\noindent {\bf Acknowledgments}\\
We acknowledge useful discussions with J.C. Winter. 
R.K.E and W.G are grateful to the Center for Theoretical Studies, ETH,
Z\"urich for hospitality during the initial stage of this work. 
The research of K.M. is supported by the startup package provided by 
Johns Hopkins University. 
G.Z. is supported by the British Science and Technology Facilities Council.
Fermilab is operated by Fermi Research Alliance, LLC under Contract
No. DE-AC02-07CH11359 with the United States Department of Energy.

\newpage

\appendix

\section{Numerical results}
\label{app:numres}

In this Appendix we present the numerical results for various
primitive amplitudes that we employ for the computation of $W+3$ jet
processes. Numerical results are presented for the phase-space point
given in Eq.(\ref{kinpoint}) for various
helicities of the external particles. For convenience, we present the
results for the ratio of a one-loop primitive amplitude and the
corresponding primitive tree-level amplitude defined
as~\footnote{Note that $A^{{\rm tree}}(1,2,3,4,5,6,7)$ denote here {\it
primitive} tree-level amplitudes, as opposed to the color ordered
amplitudes 
$A_n^\tree(1_{\bar{q}},2_q;{\sigma(3)}_g,\ldots,{\sigma(n)}_g)$
appearing e.g. eq.~(\ref{eq:atree}), where the quarks are separated by
a semicolon from the gluons.}
\be
r_L^{[j]}(1,2,3,4,5,6,7) = \frac{1}{c_\Gamma} 
\frac{A_L^{[j]}(1,2,3,4,5,6,7)}{
A^{{\rm tree}}(1,2,3,4,5,6,7)},\;\;\;
c_{\Gamma} = \frac{\Gamma(1+\epsilon)\Gamma(1-\epsilon)^2}
{(4\pi)^{2-\epsilon} \Gamma(1-2\epsilon)}, 
\ee
where in this appendix we always indicate explicitly the dependence
on the lepton momenta from the W decay.

\subsection{Numerical results for $0 \to \qqgggW$ amplitudes}

\begin{table}[h]
\begin{center}
\begin{tabular}{|l|c|c|c|}
\hline
Helicity& $\; 1/\e^2 \;$ & $1/\e$ & $\e^0$ \\
\hline
$\;A^{\tree}({1_\qb^+}\,{2_q^-}\,{3_g^+}\,{4_g^+}\,{5_g^+}\,{6_{\lb_{\vphantom{A}}}^+}\,{7_{l}^-})$&& 
 & $\;   -0.006873+i\,    0.011728\;$\\
$\;r_L^{[1]}({1_\qb^+}\,{2_q^-}\,{3_g^+}\,{4_g^+}\,{5_g^+}\,{6_{\lb_{\vphantom{A}}}^+}\,{7_{l}^-})$
&$\;-4.00000\;$
& $\;  -10.439578-i\,    9.424778$ & $\;    5.993700-i\,   19.646278\;$\\
\hline 
$\;A^{\tree}({1_\qb^+}\,{2_q^-}\,{3_g^+}\,{4_g^+}\,{5_g^-}\,{6_{\lb_{\vphantom{A}}}^+}\,{7_{l}^-})$
&& 
 & $\;    0.010248-i\,    0.007726\;$\\
$\;r_L^{[1]}({1_\qb^+}\,{2_q^-}\,{3_g^+}\,{4_g^+}\,{5_g^-}\,{6_{\lb_{\vphantom{A}}}^+}\,{7_{l}^-})$
&$\;-4.00000\;$
 & $\;  -10.439578-i\,    9.424778$ & $\;  -14.377555-i\,   37.219716\;$\\
\hline 
$\;A^{\tree}({1_\qb^+}\,{2_q^-}\,{3_g^-}\,{4_g^+}\,{5_g^+}\,{6_{\lb_{\vphantom{A}}}^+}\,{7_{l}^-})$
& &
 & $\;    0.495774-i\,    1.274796\;$\\
$\;r_L^{[1]}({1_\qb^+}\,{2_q^-}\,{3_g^-}\,{4_g^+}\,{5_g^+}\,{6_{\lb_{\vphantom{A}}}^+}\,{7_{l}^-})$
&$\;-4.00000\;$
 & $\;  -10.439578-i\,    9.424778$ & $\;   -1.039489-i\,   30.210418\;$\\
\hline 
$\;A^{\tree}({1_\qb^+}\,{2_q^-}\,{3_g^-}\,{4_g^+}\,{5_g^-}\,{6_{\lb_{\vphantom{A}}}^+}\,{7_{l}^-})$
&&
 & $\;   -0.294256-i\,    0.223277\;$\\
$\;r_L^{[1]}({1_\qb^+}\,{2_q^-}\,{3_g^-}\,{4_g^+}\,{5_g^-}\,{6_{\lb_{\vphantom{A}}}^+}\,{7_{l}^-})$
&$\;-4.00000\;$
 & $\;  -10.439578-i\,    9.424778$ & $\;   -1.444709-i\,   26.101951\;$\\
\hline
\end{tabular}
\end{center}
\caption{
The primitive tree-level amplitude
$A^{\tree}(1_\qb,2_q,3_g,4_g,5_g,6_\lb,7_l)$ and the ratio
$r_L^{[1]}(1_\qb,2_q,3_g,4_g,5_g,6_\lb,7_l)$ of a primitive one-loop
amplitude to the primitive tree-level amplitudes for various
helicities.  }
\label{table:nterm3}
\end{table}
\begin{table}[h]
\begin{center}
\begin{tabular}{|l|c|c|c|}
\hline
Helicity&$\;1/\e^2\;$&$1/\e$&$\e^0$\\
\hline
$\;A^{\tree}({1_\qb^+}\,{3_g^+}\,{2_q^-}\,{4_g^+}\,{5_g^+}\,{6_{\lb_{\vphantom{A}}}^+}\,{7_{l}^-})$&&
 & $\;   -0.005446+i\,    0.009804\;$\\
$\;r_L^{[1]}({1_\qb^+}\,{3_g^+}\,{2_q^-}\,{4_g^+}\,{5_g^+}\,{6_{\lb_{\vphantom{A}}}^+}\,{7_{l}^-})$
&$\;-3.00000\;$
 & $\;   -8.676830-i\,    6.283185$ & $\;   -1.423339-i\,   14.443863\;$\\
\hline
$\;A^{\tree}({1_\qb^+}\,{3_g^+}\,{2_q^-}\,{4_g^+}\,{5_g^-}\,{6_{\lb_{\vphantom{A}}}^+}\,{7_{l}^-})$&&
 & $\;    0.000364+i\,    0.004550\;$\\
$\;r_L^{[1]}({1_\qb^+}\,{3_g^+}\,{2_q^-}\,{4_g^+}\,{5_g^-}\,{6_{\lb_{\vphantom{A}}}^+}\,{7_{l}^-})$
&$\;-3.00000\;$
 & $\;   -8.676830-i\,    6.283185$ & $\;  -11.406265-i\,   16.485295\;$\\
\hline
$\;A^{\tree}({1_\qb^+}\,{3_g^-}\,{2_q^-}\,{4_g^+}\,{5_g^+}\,{6_{\lb_{\vphantom{A}}}^+}\,{7_{l}^-})$&&
 & $\;    0.341643-i\,    0.310960\;$\\
$\;r_L^{[1]}({1_\qb^+}\,{3_g^-}\,{2_q^-}\,{4_g^+}\,{5_g^+}\,{6_{\lb_{\vphantom{A}}}^+}\,{7_{l}^-})$
&$\;-3.00000\;$
 & $\;   -8.676830-i\,    6.283185$ & $\;   -5.430180-i\,   21.180247\;$\\
\hline
$\;A^{\tree}({1_\qb^+}\,{3_g^-}\,{2_q^-}\,{4_g^+}\,{5_g^-}\,{6_{\lb_{\vphantom{A}}}^+}\,{7_{l}^-})$&&
 & $\;    0.024966-i\,    0.156703\;$\\
$\; r_L^{[1]}({1_\qb^+}\,{3_g^-}\,{2_q^-}\,{4_g^+}\,{5_g^-}\,{6_{\lb_{\vphantom{A}}}^+}\,{7_{l}^-})$
&$\;-3.00000\;$
 & $\;   -8.676830-i\,    6.283185$ & $\;   -4.868668-i\,   21.036597\;$\\
\hline
\end{tabular}
\end{center}
\caption{
The primitive tree-level amplitude
$A^{\tree}(1_\qb,3_g,2_q,4_g,5_g,6_\lb,7_l)$ and the ratio
$r_L^{[1]}(1_\qb,3_g,2_q,4_g,5_g,6_\lb,7_l)$ of a primitive one-loop
amplitude to tree-level primitive amplitude for various helicities.  }
\label{table:nterm4}
\end{table}

\newpage 
\begin{table}[h]
\begin{center}
\begin{tabular}{|l|c|c|c|}
\hline
Helicity& $\; 1/\e^2 \;$ & $1/\e$ & $\e^0$ \\
\hline
$\;A^{\tree}({1_\qb^+}\,{3_g^+}\,{4_g^+}\,{2_q^-}\,{5_g^+}\,{6_{\lb_{\vphantom{A}}}^+}\,{7_{l}^-})$&&
 & $\;   -0.005563-i\,    0.030746\;$\\
$\;r_L^{[1]}({1_\qb^+}\,{3_g^+}\,{4_g^+}\,{2_q^-}\,{5_g^+}\,{6_{\lb_{\vphantom{A}}}^+}\,{7_{l}^-})$
&$\;-2.00000\;$
 & $\;   -7.835662-i\,    3.141593$ & $\;   13.662096-i\,   25.637707\;$\\
\hline
$\;A^{\tree}({1_\qb^+}\,{3_g^+}\,{4_g^+}\,{2_q^-}\,{5_g^-}\,{6_{\lb_{\vphantom{A}}}^+}\,{7_{l}^-})$&&
 & $\;    0.022677+i\,    0.085524\;$\\
$\;r_L^{[1]}({1_\qb^+}\,{3_g^+}\,{4_g^+}\,{2_q^-}\,{5_g^-}\,{6_{\lb_{\vphantom{A}}}^+}\,{7_{l}^-})$
&$\;-2.00000\;$
 & $\;   -7.835662-i\,    3.141593$ & $\;   -9.177581-i\,   16.265480\;$\\
\hline
$\;A^{\tree}({1_\qb^+}\,{3_g^-}\,{4_g^+}\,{2_q^-}\,{5_g^+}\,{6_{\lb_{\vphantom{A}}}^+}\,{7_{l}^-})$&&
 & $\;   -0.098988+i\,    1.958409\;$\\
$\;r_L^{[1]}({1_\qb^+}\,{3_g^-}\,{4_g^+}\,{2_q^-}\,{5_g^+}\,{6_{\lb_{\vphantom{A}}}^+}\,{7_{l}^-})$
&$\;-2.00000\;$
 & $\;   -7.835662-i\,    3.141593$ & $\;  -12.140461-i\,   15.924761\;$\\
\hline
$\;A^{\tree}({1_\qb^+}\,{3_g^-}\,{4_g^+}\,{2_q^-}\,{5_g^-}\,{6_{\lb_{\vphantom{A}}}^+}\,{7_{l}^-})$&&
 & $\;   -0.283565+i\,    0.841833\;$\\
$\; r_L^{[1]}({1_\qb^+}\,{3_g^-}\,{4_g^+}\,{2_q^-}\,{5_g^-}\,{6_{\lb_{\vphantom{A}}}^+}\,{7_{l}^-})$
&$\;-2.00000\;$
 & $\;   -7.835662-i\,    3.141593$ & $\;  -13.465828-i\,   13.730719\;$\\
\hline
\end{tabular}
\end{center}
\caption{
The primitive tree-level amplitude 
$A^{\tree}(1_\qb,3_g,4_g,2_q,5_g,6_\lb,7_l)$ and
the ratio $r_L^{[1]}(1_\qb,3_g,4_g,2_q,5_g,6_\lb,7_l)$
of a primitive one-loop amplitude to the primitive 
tree-level amplitude for various helicities.
}
\label{table:nterm5}
\end{table}

\begin{table}[h]
\begin{center}
\begin{tabular}{|l|c|c|c|}
\hline
Helicity&$\;1/\e^2\;$&$1/\e$&$\e^0$\\
\hline
$\;A^{\tree}({1_\qb^+}\,{3_g^+}\,{4_g^+}\,{5_g^+}\,{2_q^-}\,{6_{\lb_{\vphantom{A}}}^+}\,{7_{l}^-})$&&
 & $\;    0.017883+i\,    0.009214\;$\\
$\;r_L^{[1]}({1_\qb^+}\,{3_g^+}\,{4_g^+}\,{5_g^+}\,{2_q^-}\,{6_{\lb_{\vphantom{A}}}^+}\,{7_{l}^-})$
&$\;-1.00000\;$
 & $\;   -3.334232-i\,    0.000000$ & $\;   11.973924-i\,    8.033958\;$\\
\hline
$\;A^{\tree}({1_\qb^+}\,{3_g^+}\,{4_g^+}\,{5_g^-}\,{2_q^-}\,{6_{\lb_{\vphantom{A}}}^+}\,{7_{l}^-})$&&
 & $\;   -0.033289-i\,    0.082348\;$\\
$\;r_L^{[1]}({1_\qb^+}\,{3_g^+}\,{4_g^+}\,{5_g^-}\,{2_q^-}\,{6_{\lb_{\vphantom{A}}}^+}\,{7_{l}^-})$
&$\;-1.00000\;$
 & $\;   -3.334232+i\,    0.000000$ & $\;   -1.783190+i\,    1.552944\;$\\
\hline
$\;A^{\tree}({1_\qb^+}\,{3_g^-}\,{4_g^+}\,{5_g^+}\,{2_q^-}\,{6_{\lb_{\vphantom{A}}}^+}\,{7_{l}^-})$&&
 & $\;   -0.738428-i\,    0.372652\;$\\
$\;r_L^{[1]}({1_\qb^+}\,{3_g^-}\,{4_g^+}\,{5_g^+}\,{2_q^-}\,{6_{\lb_{\vphantom{A}}}^+}\,{7_{l}^-})$
&$\;-1.00000\;$
 & $\;   -3.334232-i\,    0.000000$ & $\;   -5.654597+i\,    0.276608\;$\\
\hline
$\;A^{\tree}({1_\qb^+}\,{3_g^-}\,{4_g^+}\,{5_g^-}\,{2_q^-}\,{6_{\lb_{\vphantom{A}}}^+}\,{7_{l}^-})$&&
 & $\;    0.552856-i\,    0.461853\;$\\
$\;r_L^{[1]}({1_\qb^+}\,{3_g^-}\,{4_g^+}\,{5_g^-}\,{2_q^-}\,{6_{\lb_{\vphantom{A}}}^+}\,{7_{l}^-})$
&$\;-1.00000\;$
 & $\;   -3.334232+i\,    0.000000$ & $\;   -6.461431+i\,    1.451815\;$\\
\hline
\end{tabular}
\end{center}
\caption{
The primitive tree-level amplitude 
$A^{\tree}(1_\qb,3_g,4_g,5_g,2_q,6_\lb,7_l)$ and
the ratio $r_L^{[1]}(1_\qb,3_g,4_g,5_g,2_q,6_\lb,7_l)$
of a  primitive one-loop amplitude to the primitive 
tree-level amplitude 
for various helicities.}
\label{table:nterm6}
\end{table}

\newpage

\begin{table}[h]
\begin{center}
\begin{tabular}{|l|c|c|c|}
\hline
Helicity&$\;1/\e^2\;$&$1/\e$&$\e^0$\\
\hline
$\;A^{\tree}({1_\qb^+}\,{2_q^-}\,{3_g^+}\,{4_g^+}\,{5_g^+}\,{6_{\lb_{\vphantom{A}}}^+}\,{7_{l}^-})$&&
 & $\;   -0.006873+i\,    0.011728\;$\\
$\;r_L^{[1/2]}({1_\qb^+}\,{2_q^-}\,{3_g^+}\,{4_g^+}\,{5_g^+}\,{6_{\lb_{\vphantom{A}}}^+}\,{7_{l}^-})$
&$\;0.00000\;$
& $\;    0.000000-i\,    0.000000$ & $\;    -6.001512-i\,   26.601839\;$\\
\hline

$\;A^{\tree}({1_\qb^+}\,{2_q^-}\,{3_g^+}\,{4_g^+}\,{5_g^-}\,{6_{\lb_{\vphantom{A}}}^+}\,{7_{l}^-})$&&
 & $\;    0.010248-i\,    0.007726\;$\\
$\;r_L^{[1/2]}({1_\qb^+}\,{2_q^-}\,{3_g^+}\,{4_g^+}\,{5_g^-}\,{6_{\lb_{\vphantom{A}}}^+}\,{7_{l}^-})$
&$\;0.00000\;$
 & $\;    0.000000-i\,    0.000000$ & $\;    7.227836-i\,    4.090839\;$\\
\hline
$\;A^{\tree}({1_\qb^+}\,{2_q^-}\,{3_g^-}\,{4_g^+}\,{5_g^+}\,{6_{\lb_{\vphantom{A}}}^+}\,{7_{l}^-})$&&
 & $\;    0.495774-i\,    1.274796\;$\\
$\;r_L^{[1/2]}({1_\qb^+}\,{2_q^-}\,{3_g^-}\,{4_g^+}\,{5_g^+}\,{6_{\lb_{\vphantom{A}}}^+}\,{7_{l}^-})$
&$\;0.00000\;$
 & $\;    0.000000+i\,    0.000000$ & $\;    0.096288-i\,    0.114398\;$\\
\hline
$\;A^{\tree}({1_\qb^+}\,{2_q^-}\,{3_g^-}\,{4_g^+}\,{5_g^-}\,{6_{\lb_{\vphantom{A}}}^+}\,{7_{l}^-})$&&
 & $\;   -0.294256-i\,    0.223277\;$\\
$\;r_L^{[1/2]}({1_\qb^+}\,{2_q^-}\,{3_g^-}\,{4_g^+}\,{5_g^-}\,{6_{\lb_{\vphantom{A}}}^+}\,{7_{l}^-})$
&$\;0.00000\;$
 & $\;    0.000000+i\,    0.000000$ & $\;    0.164410-i\,    0.134601\;$\\
\hline
\end{tabular} \end{center} \caption{
The primitive tree-level amplitude $A^{\tree}(1_\qb,2_q,3_g,4_g,5_g,6_\lb,7_l)$ and
the ratio $r_L^{[1/2]}(1_\qb,2_q,3_g,4_g,5_g,6_\lb,7_l)$
of a  primitive one-loop amplitude to the primitive 
tree-level amplitude 
for various helicities.
}
\label{table:nf1}
\end{table}

\begin{table}[h]
\begin{center}
\begin{tabular}{|l|c|c|c|}
\hline
Helicity&$\;1/\e^2\;$&$1/\e$&$\e^0$\\
\hline
$\;A^{\tree}({1_\qb^+}\,{3_g^+}\,{2_q^-}\,{4_g^+}\,{5_g^+}\,{6_{\lb_{\vphantom{A}}}^+}\,{7_{l}^-})$&&
 & $\;   -0.005446+i\,    0.009804\;$\\
$\;r_L^{[1/2]}({1_\qb^+}\,{3_g^+}\,{2_q^-}\,{4_g^+}\,{5_g^+}\,{6_{\lb_{\vphantom{A}}}^+}\,{7_{l}^-})$
&$\;0.00000\;$
 & $\;    0.000000-i\,    0.000000$ & $\;    -0.127241-i\,    1.316987\;$\\
\hline
$\;A^{\tree}({1_\qb^+}\,{3_g^+}\,{2_q^-}\,{4_g^+}\,{5_g^-}\,{6_{\lb_{\vphantom{A}}}^+}\,{7_{l}^-})$&&
 & $\;    0.000364+i\,    0.004550\;$\\
$\;r_L^{[1/2]}({1_\qb^+}\,{3_g^+}\,{2_q^-}\,{4_g^+}\,{5_g^-}\,{6_{\lb_{\vphantom{A}}}^+}\,{7_{l}^-})$
&$\;0.00000\;$
 & $\;    0.000000+i\,    0.000000$ & $\;    0.000000+i\,    0.000000\;$\\
\hline
$\;A^{\tree}({1_\qb^+}\,{3_g^-}\,{2_q^-}\,{4_g^+}\,{5_g^+}\,{6_{\lb_{\vphantom{A}}}^+}\,{7_{l}^-})$&&
 & $\;    0.341643-i\,    0.310960\;$\\
$\;r_L^{[1/2]}({1_\qb^+}\,{3_g^-}\,{2_q^-}\,{4_g^+}\,{5_g^+}\,{6_{\lb_{\vphantom{A}}}^+}\,{7_{l}^-})$
&$\;0.00000\;$
 & $\;    0.000000+i\,    0.000000$ & $\;    -0.020783-i\,    0.001593\;$\\
\hline
$\;A^{\tree}({1_\qb^+}\,{3_g^-}\,{2_q^-}\,{4_g^+}\,{5_g^-}\,{6_{\lb_{\vphantom{A}}}^+}\,{7_{l}^-})$&&
 & $\;    0.024966-i\,    0.156703\;$\\
$\;r_L^{[1/2]}({1_\qb^+}\,{3_g^-}\,{2_q^-}\,{4_g^+}\,{5_g^-}\,{6_{\lb_{\vphantom{A}}}^+}\,{7_{l}^-})$
&$\;0.00000\;$
 & $\;    0.000000-i\,    0.000000$ & $\;    0.000000-i\,    0.000000\;$\\
\hline
\end{tabular} \end{center} \caption{The primitive tree-level
amplitude $A^{\tree}(1_\qb,3_g,2_q,4_g,5_g,6_\lb,7_l)$ and the ratio
$r_L^{[1/2]}(1_\qb,3_g,2_q,4_g,5_g,6_\lb,7_l)$ of the primitive
one-loop amplitude to the primitive tree-level amplitude for various
helicities.
}
\label{table:nf2}
\end{table}

\clearpage

\newpage

\subsection{Numerical results for $0 \to \qqQQgW$ amplitudes}

Numerical results for amplitudes with two quark pairs, a $W$ boson and a gluon 
are presented in Tables below.

\begin{table}[h]
\begin{center}
\begin{tabular}{|l|c|c|c|}
\hline
Helicity&$\;1/\e^2\;$&$1/\e$&$\e^0$\\
\hline
$\;A^{\tree}({1_\qb^+}\,{5_g^+}\,{2_q^-}\,{3_\Qb^-}\,\,{4_Q^+}\,{6_{\lb_{\vphantom{A}}}^+}\,{7_{l}^-})$&&
 & $\;   -1.347977-i\,    0.593626\;$\\
$\;r_{L}^{[1],a}({1_\qb^+}\,{5_g^+}\,{2_q^-}\,{3_\Qb^-}\,{4_Q^+}\,{6_{\lb_{\vphantom{A}}}^+}\,{7_{l}^-})$
&$\;-2.00000\;$
& $\;   -1.906388-i\,    6.283185$ & $\;   12.513239-i\,   16.727811\;$\\
\hline
$\;A^{\tree}({1_\qb^+}\,{5_g^-}\,{2_q^-}\,{3_\Qb^-}\,\,{4_Q^+}\,{6_{\lb_{\vphantom{A}}}^+}\,{7_{l}^-})$&&
 & $\;   -0.570749-i\,    0.316836\;$\\
$\;r_{L}^{[1],a}({1_\qb^+}\,{5_g^-}\,{2_q^-}\,{3_\Qb^-}\,{4_Q^+}\,{6_{\lb_{\vphantom{A}}}^+}\,{7_{l}^-})$
&$\;-2.00000\;$
& $\;   -1.906388-i\,    6.283185$ & $\;   12.452841-i\,   14.482266\;$\\
\hline 
$\;A^{\tree}({1_\qb^+}\,{2_q^-}\,{5_g^+}\,{3_\Qb^-}\,{4_Q^+}\,{6_{\lb_{\vphantom{A}}}^+}\,{7_{l}^-})$&&
 & $\;    0.734834-i\,    0.360895\;$\\
$\;r_{L}^{[1],a}({1_\qb^+}\,{2_q^-}\,{5_g^+}\,{3_\Qb^-}\,{4_Q^+}\,{6_{\lb_{\vphantom{A}}}^+}\,{7_{l}^-})$
&$\;-3.00000\;$
& $\;   -6.083408-i\,    3.141593$ & $\;    9.466663-i\,    4.461718\;$\\
\hline 
$\;A^{\tree}({1_\qb^+}\,{2_q^-}\,{5_g^-}\,{3_\Qb^-}\,{4_Q^+}\,{6_{\lb_{\vphantom{A}}}^+}\,{7_{l}^-})$&&
 & $\;    0.421057+i\,    0.463392\;$\\
$\;r_{L}^{[1],a}({1_\qb^+}\,{2_q^-}\,{5_g^-}\,{3_\Qb^-}\,{4_Q^+}\,{6_{\lb_{\vphantom{A}}}^+}\,{7_{l}^-})$
&$\;-3.00000\;$
& $\;   -6.083408-i\,    3.141593$ & $\;    7.337316-i\,    3.094966\;$\\
\hline
$\;A^{\tree}({1_\qb^+}\,{2_q^-}\,{3_\Qb^-}\,{5_g^+}\,{4_Q^+}\,{6_{\lb_{\vphantom{A}}}^+}\,{7_{l}^-})$&&
 & $\;   -0.288607-i\,    0.043034\;$\\
$\;r_{L}^{[1],a}({1_\qb^+}\,{2_q^-}\,{3_\Qb^-}\,{5_g^+}\,{4_Q^+}\,{6_{\lb_{\vphantom{A}}}^+}\,{7_{l}^-})$
&$\;-2.00000\;$
& $\;   -1.906388-i\,    6.283185$ & $\;   14.470591-i\,   16.520704\;$\\
\hline
$\;A^{\tree}({1_\qb^+}\,{2_q^-}\,{3_\Qb^-}\,{5_g^-}\,{4_Q^+}\,{6_{\lb_{\vphantom{A}}}^+}\,{7_{l}^-})$&&
 & $\;   -0.148841+i\,    0.001847\;$\\
$\;r_{L}^{[1],a}({1_\qb^+}\,{2_q^-}\,{3_\Qb^-}\,{5_g^-}\,{4_Q^+}\,{6_{\lb_{\vphantom{A}}}^+}\,{7_{l}^-})$
&$\;-2.00000\;$
 & $\;   -1.906388-i\,    6.283185$ & $\;   10.478262-i\,   16.315474\;$\\
\hline
$\;A^{\tree}({1_\qb^+}\,{2_q^-}\,{3_\Qb^-}\,{4_Q^+}\,{5_g^+}\,{6_{\lb_{\vphantom{A}}}^+}\,{7_{l}^-})$&&
 & $\;    0.901749+i\,    0.997556\;$\\
$\;r_{L}^{[1],a}({1_\qb^+}\,{2_q^-}\,{3_\Qb^-}\,{4_Q^+}\,{5_g^+}\,{6_{\lb_{\vphantom{A}}}^+}\,{7_{l}^-})$
&$\;-3.00000\;$
& $\;   -5.946351-i\,    9.424778$ & $\;   10.012255-i\,   29.539947\;$\\
\hline
$\;A^{\tree}({1_\qb^+}\,{2_q^-}\,{3_\Qb^-}\,{4_Q^+}\,{5_g^-}\,{6_{\lb_{\vphantom{A}}}^+}\,{7_{l}^-})$&&
& $\;    0.298533-i\,    0.147741\;$\\
$\;r_{L}^{[1],a}({1_\qb^+}\,{2_q^-}\,{3_\Qb^-}\,{4_Q^+}\,{5_g^-}\,{6_{\lb_{\vphantom{A}}}^+}\,{7_{l}^-})$
&$\;-3.00000\;$
 & $\;   -5.946351-i\,    9.424778$ & $\;    9.462997-i\,   25.465941\;$\\
\hline
\end{tabular} \end{center} \caption{
The primitive tree-level four-quark amplitudes $A^{\tree}$ and 
the ratio $r_L^{[1],a}$ of the one-loop class $a$  primitive amplitude  to the
corresponding primitive tree-level amplitude 
for various helicities.
}
\label{tab7}
\end{table}

\begin{table}[h]
\begin{center}
\begin{tabular}{|l|c|c|c|}
\hline
Helicity&$\;1/\e^2\;$&$1/\e$&$\e^0$\\
\hline
$    \;A^{\tree}({1_\qb^+}\,{5_g^+}\,{2_q^-}\,{4_\Qb^+}\,\,{3_Q^-}\,{6_{\lb_{\vphantom{A}}}^+}\,{7_{l}^-})$&&
 & $\;   -1.347977-i\,    0.593626\;$\\
$\;r_{L}^{[1],a}({1_\qb^+}\,{5_g^+}\,{2_q^-}\,{4_\Qb^+}\,{3_Q^-}\,{6_{\lb_{\vphantom{A}}}^+}\,{7_{l}^-})$
&$\;-2.00000\;$
 & $\;   -3.109446+i\,    0.000000$ & $\;    7.650907-i\,    1.630983\;$\\
\hline
$    \;A^{\tree}({1_\qb^+}\,{5_g^-}\,{2_q^-}\,{4_\Qb^+}\,\,{3_Q^-}\,{6_{\lb_{\vphantom{A}}}^+}\,{7_{l}^-})$&&
 & $\;   -0.570749-i\,    0.316836\;$\\
$\;r_{L}^{[1],a}({1_\qb^+}\,{5_g^-}\,{2_q^-}\,{4_\Qb^+}\,{3_Q^-}\,{6_{\lb_{\vphantom{A}}}^+}\,{7_{l}^-})$
&$\;-2.00000\;$
 & $\;   -3.109446+i\,    0.000000$ & $\;    7.388565+i\,    2.000057\;$\\
\hline 
$\;    A^{\tree}({1_\qb^+}\,{2_q^-}\,{5_g^+}\,{4_\Qb^+}\,{3_Q^-}\,{6_{\lb_{\vphantom{A}}}^+}\,{7_{l}^-})$&&
 & $\;    0.446227-i\,    0.403929\;$\\
$\;r_{L}^{[1],a}({1_\qb^+}\,{2_q^-}\,{5_g^+}\,{4_\Qb^+}\,{3_Q^-}\,{6_{\lb_{\vphantom{A}}}^+}\,{7_{l}^-})$
&$\;-3.00000\;$
 & $\;   -6.730261-i\,    3.141593$ & $\;    8.835307-i\,   10.218599\;$\\
\hline 
$\;    A^{\tree}({1_\qb^+}\,{2_q^-}\,{5_g^-}\,{4_\Qb^+}\,{3_Q^-}\,{6_{\lb_{\vphantom{A}}}^+}\,{7_{l}^-})$&&
 & $\;    0.272217+i\,    0.464577\;$\\
$\;r_{L}^{[1],a}({1_\qb^+}\,{2_q^-}\,{5_g^-}\,{4_\Qb^+}\,{3_Q^-}\,{6_{\lb_{\vphantom{A}}}^+}\,{7_{l}^-})$
&$\;-3.00000\;$
 & $\;   -6.730261-i\,    3.141593$ & $\;    2.417473-i\,   12.686072\;$\\
\hline
$\;    A^{\tree}({1_\qb^+}\,{2_q^-}\,{4_\Qb^+}\,{5_g^+}\,{3_Q^-}\,{6_{\lb_{\vphantom{A}}}^+}\,{7_{l}^-})$&&
 & $\;    0.288607+i\,    0.043034\;$\\
$\;r_{L}^{[1],a}({1_\qb^+}\,{2_q^-}\,{4_\Qb^+}\,{5_g^+}\,{3_Q^-}\,{6_{\lb_{\vphantom{A}}}^+}\,{7_{l}^-})$
&$\;-2.00000\;$
 & $\;   -3.109446+i\,    0.000000$ & $\;   11.636607+i\,    2.670357\;$\\
\hline
$\;    A^{\tree}({1_\qb^+}\,{2_q^-}\,{4_\Qb^+}\,{5_g^-}\,{3_Q^-}\,{6_{\lb_{\vphantom{A}}}^+}\,{7_{l}^-})$&&
 & $\;    0.148841-i\,    0.001185\;$\\
$\;r_{L}^{[1],a}({1_\qb^+}\,{2_q^-}\,{4_\Qb^+}\,{5_g^-}\,{3_Q^-}\,{6_{\lb_{\vphantom{A}}}^+}\,{7_{l}^-})$
&$\;-2.00000\;$
& $\;   -3.109446+i\,    0.000000$ & $\;    3.098383-i\,    0.972210\;$\\
\hline
$\;    A^{\tree}({1_\qb^+}\,{2_q^-}\,{4_\Qb^+}\,{3_Q^-}\,{5_g^+}\,{6_{\lb_{\vphantom{A}}}^+}\,{7_{l}^-})$&&
 & $\;    0.613143+i\,    0.954522\;$\\
$\;r_{L}^{[1],a}({1_\qb^+}\,{2_q^-}\,{4_\Qb^+}\,{3_Q^-}\,{5_g^+}\,{6_{\lb_{\vphantom{A}}}^+}\,{7_{l}^-})$
&$\;-3.00000\;$
 & $\;   -6.502556-i\,    3.141593$ & $\;    9.563123-i\,   12.742650\;$\\
\hline
$\;    A^{\tree}({1_\qb^+}\,{2_q^-}\,{4_\Qb^+}\,{3_Q^-}\,{5_g^-}\,{6_{\lb_{\vphantom{A}}}^+}\,{7_{l}^-})$&&
 & $\;    0.149692-i\,    0.146556\;$\\
$\;r_{L}^{[1],a}({1_\qb^+}\,{2_q^-}\,{4_\Qb^+}\,{3_Q^-}\,{5_g^-}\,{6_{\lb_{\vphantom{A}}}^+}\,{7_{l}^-})$
&$\;-3.00000\;$
 & $\;   -6.502556-i\,    3.141593$ & $\;    7.832425-i\,    8.905177\;$\\
\hline
\end{tabular} \end{center} \caption{
The primitive  tree-level four-quark amplitudes $A^{\tree}$ and 
the ratio $r_L^{[1],a}$ of the   primitive one-loop class $a$ amplitude  to the
corresponding  primitive tree-level amplitude 
for various helicities. Note 
 the different ordering of the quarks
$Q$ and $\Qb$ compared  to Table~7. 
}
\label{tab7a}
\end{table}

\begin{table}[h]
\begin{center}
\begin{tabular}{|l|c|c|c|}
\hline
Helicity&$\;1/\e^2\;$&$1/\e$&$\e^0$\\
\hline
$\;A^{\tree}({1_\qb^+}\,{5_g^+}\,{4_Q^+}\,{3_\Qb^-}\,{2_q^-}\,{6_{\lb_{\vphantom{A}}}^+}\,{7_{l}^-})$&&
 & $\;   -0.901749-i\;    0.997556\;$\\
$\;r_L^{[1],b}({1_\qb^+}\,{5_g^+}\,{4_Q^+}\,{3_\Qb^-}\,{2_q^-}\,{6_{\lb_{\vphantom{A}}}^+}\,{7_{l}^-})$
&$\;-1.00000\;$
 & $\;   -3.334232+i\,    0.000000$ & $\;   -8.027321-i\,    0.968722\;$\\
\hline
$\;A^{\tree}({1_\qb^+}\,{5_g^-}\,{4_Q^+}\,{3_\Qb^-}\,{2_q^-}\,{6_{\lb_{\vphantom{A}}}^+}\,{7_{l}^-})$&&
 & $\;   -0.298533+i\,    0.147741\;$\\
$\;r_L^{[1],b}({1_\qb^+}\,{5_g^-}\,{4_Q^+}\,{3_\Qb^-}\,{2_q^-}\,{6_{\lb_{\vphantom{A}}}^+}\,{7_{l}^-})$
&$\;-1.00000\;$
& $\;   -3.334232+i\,    0.000000$ & $\;   -4.397007-i\,    3.109084\;$\\
\hline
$\;A^{\tree}({1_\qb^+}\,{4_Q^+}\,{{5_g^+}\,3_\Qb^-}\,{2_q^-}\,{6_{\lb_{\vphantom{A}}}^+}\,{7_{l}^-})$&&
 & $\;    0.288607+i\,    0.043034\;$\\
$\;r_L^{[1],b}({1_\qb^+}\,{4_Q^+}\,{{5_g^+}\,3_\Qb^-}\,{2_q^-}\,{6_{\lb_{\vphantom{A}}}^+}\,{7_{l}^-})$
&$\;-1.00000\;$
 & $\;   -3.334232+i\,    0.000000$ & $\;   -5.997531-i\,    0.856042\;$\\
\hline
$\;A^{\tree}({1_\qb^+}\,{4_Q^+}\,{5_g^-}\,{3_\Qb^-}\,{2_q^-}\,{6_{\lb_{\vphantom{A}}}^+}\,{7_{l}^-})$&&
 & $\;    0.148841-i\,    0.001185\;$\\
$\;r_L^{[1],b}({1_\qb^+}\,{4_Q^+}\,{5_g^-}\,{3_\Qb^-}\,{2_q^-}\,{6_{\lb_{\vphantom{A}}}^+}\,{7_{l}^-})$
&$\;-1.00000\;$
 & $\;   -3.334232+i\,    0.000000$ & $\;   -6.462353-i\,    0.674696\;$\\
\hline
$\;A^{\tree}({1_\qb^+}\,{4_Q^+}\,{3_\Qb^-}\,{5_g^+}\,{2_q^-}\,{6_{\lb_{\vphantom{A}}}^+}\,{7_{l}^-})$&&
 & $\;   -0.734834+i\,    0.360895\;$\\
$\;r_L^{[1],b}({1_\qb^+}\,{4_Q^+}\,{3_\Qb^-}\,{5_g^+}\,{2_q^-}\,{6_{\lb_{\vphantom{A}}}^+}\,{7_{l}^-})$
&$\;-1.00000\;$
 & $\;   -3.334232-i\,    0.000000$ & $\;   -6.039221-i\,    0.200173\;$\\
\hline
$\;A^{\tree}({1_\qb^+}\,{4_Q^+}\,{3_\Qb^-}\,{5_g^-}\,{2_q^-}\,{6_{\lb_{\vphantom{A}}}^+}\,{7_{l}^-})$&&
 & $\;   -0.421057-i\,    0.463392\;$\\
$\;r_L^{[1],b}({1_\qb^+}\,{4_Q^+}\,{3_\Qb^-}\,{5_g^-}\,{2_q^-}\,{6_{\lb_{\vphantom{A}}}^+}\,{7_{l}^-})$
&$\;-1.00000\;$
 & $\;   -3.334232+i\,    0.000000$ & $\;   -6.632788-i\,    0.178166\;$\\
\hline
$\;A^{\tree}({1_\qb^+}\,{3_Q^-}\,{4_\Qb^+}\,{2_q^-}\,{5_g^+}\,{6_{\lb_{\vphantom{A}}}^+}\,{7_{l}^-})$&&
 & $\;    1.347977+i\,    0.593626\;$\\
$\;r_L^{[1],b}({1_\qb^+}\,{3_Q^-}\,{4_\Qb^+}\,{2_q^-}\,{5_g^+}\,{6_{\lb_{\vphantom{A}}}^+}\,{7_{l}^-})$
&$\;-2.00000\;$
 & $\;   -7.835662-i\,    3.141593$ & $\;  -12.743438-i\,   16.092839\;$\\
\hline
$\;A^{\tree}({1_\qb^+}\,{3_Q^-}\,{4_\Qb^+}\,{2_q^-}\,{5_g^-}\,{6_{\lb_{\vphantom{A}}}^+}\,{7_{l}^-})$&&
 & $\;    0.570749+i\,    0.316836\;$\\
$\;r_L^{[1],b}({1_\qb^+}\,{3_Q^-}\,{4_\Qb^+}\,{2_q^-}\,{5_g^-}\,{6_{\lb_{\vphantom{A}}}^+}\,{7_{l}^-})$
&$\;-2.00000\;$
 & $\;   -7.835662-i\,    3.141593$ & $\;  -14.463179-i\,   14.071065\;$\\
\hline
\end{tabular} \end{center} \caption{
The  primitive tree-level four-quark amplitudes $A^{\tree}$ and 
the ratio $r_L^{[1],b}$
of the one-loop class $b$  primitive amplitude  to the
corresponding primitive tree-level amplitude 
for various helicities.
}
\label{tab8}
\end{table}

\begin{table}[h]
\begin{center}
\begin{tabular}{|l|c|c|c|}
\hline
Helicity&$\;1/\e^2\;$&$1/\e$&$\e^0$\\
\hline
$\;A^{\tree}({1_\qb^+}\,{5_g^+}\,{4_Q^+}\,{3_\Qb^-}\,{2_q^-}\,{6_{\lb_{\vphantom{A}}}^+}\,{7_{l}^-})$&&
 & $\;   -0.901749-i\,    0.997556\;$\\
$\;r_L^{[1],c}({1_\qb^+}\,{5_g^+}\,{4_Q^+}\,{3_\Qb^-}\,{2_q^-}\,{6_{\lb_{\vphantom{A}}}^+}\,{7_{l}^-})$
&$\;-1.00000\;$
 & $\;   -3.826559-i\,    0.000000$ & $\;  -10.068901-i\,    0.284089\;$\\
\hline
$\;A^{\tree}({1_\qb^+}\,{5_g^-}\,{4_Q^+}\,{3_\Qb^-}\,{2_q^-}\,{6_{\lb_{\vphantom{A}}}^+}\,{7_{l}^-})$&&
 & $\;   -0.298533+i\,    0.147741\;$\\
$\;r_L^{[1],c}({1_\qb^+}\,{5_g^-}\,{4_Q^+}\,{3_\Qb^-}\,{2_q^-}\,{6_{\lb_{\vphantom{A}}}^+}\,{7_{l}^-})$
&$\;-1.00000\;$
 & $\;   -3.826559-i\,    0.000000$ & $\;   -9.118256-i\,    0.316181\;$\\
\hline
$\;A^{\tree}({1_\qb^+}\,{4_Q^+}\,{{5_g^-}\,3_\Qb^-}\,{2_q^-}\,{6_{\lb_{\vphantom{A}}}^+}\,{7_{l}^-})$&&
 & $\;    0.288607+i\,    0.043034\;$\\
$\;r_L^{[1],c}({1_\qb^+}\,{4_Q^+}\,{{5_g^-}\,3_\Qb^-}\,{2_q^-}\,{6_{\lb_{\vphantom{A}}}^+}\,{7_{l}^-})$
&$\;-2.00000\;$
 & $\;   -5.954375-i\,    3.141593$ & $\;   -6.343258-i\,    9.293058\;$\\
\hline
$\;A^{\tree}({1_\qb^+}\,{4_Q^+}\,{5_g^-}\,{3_\Qb^-}\,{2_q^-}\,{6_{\lb_{\vphantom{A}}}^+}\,{7_{l}^-})$&&
 & $\;    0.148841-i\,    0.001185\;$\\
$\;r_L^{[1],c}({1_\qb^+}\,{4_Q^+}\,{5_g^-}\,{3_\Qb^-}\,{2_q^-}\,{6_{\lb_{\vphantom{A}}}^+}\,{7_{l}^-})$
&$\;-2.00000\;$
 & $\;   -5.954375-i\,    3.141593$ & $\;   -8.567533-i\,   11.440611\;$\\
\hline
$\;A^{\tree}({1_\qb^+}\,{4_Q^+}\,{3_\Qb^-}\,{5_g^+}\,{2_q^-}\,{6_{\lb_{\vphantom{A}}}^+}\,{7_{l}^-})$&&
 & $\;   -0.734834+i\,    0.360895\;$\\
$\;r_L^{[1],c}({1_\qb^+}\,{4_Q^+}\,{3_\Qb^-}\,{5_g^+}\,{2_q^-}\,{6_{\lb_{\vphantom{A}}}^+}\,{7_{l}^-})$
&$\;-1.00000\;$
 & $\;   -3.826559+i\,    0.000000$ & $\;   -9.495931+i\,    0.100657\;$\\
\hline
$\;A^{\tree}({1_\qb^+}\,{4_Q^+}\,{3_\Qb^-}\,{5_g^-}\,{2_q^-}\,{6_{\lb_{\vphantom{A}}}^+}\,{7_{l}^-})$&&
 & $\;   -0.421057-i\,    0.463392\;$\\
$\;r_L^{[1],c}({1_\qb^+}\,{4_Q^+}\,{3_\Qb^-}\,{5_g^-}\,{2_q^-}\,{6_{\lb_{\vphantom{A}}}^+}\,{7_{l}^-})$
&$\;-1.00000\;$
 & $\;   -3.826559-i\,    0.000000$ & $\;   -9.538700-i\,    0.299195\;$\\
\hline
$\;A^{\tree}({1_\qb^+}\,{3_Q^-}\,{4_\Qb^+}\,{2_q^-}\,{5_g^+}\,{6_{\lb_{\vphantom{A}}}^+}\,{7_{l}^-})$&&
 & $\;    1.347977+i\,    0.593626\;$\\
$\;r_L^{[1],c}({1_\qb^+}\,{3_Q^-}\,{4_\Qb^+}\,{2_q^-}\,{5_g^+}\,{6_{\lb_{\vphantom{A}}}^+}\,{7_{l}^-})$
&$\;-1.00000\;$
 & $\;   -3.826559+i\,    0.000000$ & $\;   -9.696279+i\,    0.000000\;$\\
\hline
$\;A^{\tree}({1_\qb^+}\,{3_Q^-}\,{4_\Qb^+}\,{2_q^-}\,{5_g^-}\,{6_{\lb_{\vphantom{A}}}^+}\,{7_{l}^-})$&&
 & $\;    0.570749+i\,    0.316836\;$\\
$\;r_L^{[1],c}({1_\qb^+}\,{3_Q^-}\,{4_\Qb^+}\,{2_q^-}\,{5_g^-}\,{6_{\lb_{\vphantom{A}}}^+}\,{7_{l}^-})$
&$\;-1.00000\;$
 & $\;   -3.826559+i\,    0.000000$ & $\;   -9.696279+i\,    0.000000\;$\\
\hline
\end{tabular} \end{center} \caption{
The  primitive tree-level four-quark amplitudes $A^{\tree}$ and 
the ratio $r_L^{[1],c}$ 
of the primitive one-loop class $c$ amplitude  to the
corresponding primitive tree-level amplitude 
for various helicities.
}
\label{tab9}
\end{table}

\begin{table}[h]
\begin{center}
\begin{tabular}{|l|c|c|c|}
\hline
Helicity&$\;1/\e^2\;$&$1/\e$&$\e^0$\\
\hline
$\;A^{\tree}({1_\qb^+}\,{5_g^+}\,{4_Q^+}\,{3_\Qb^-}\,{2_q^-}\,{6_{\lb_{\vphantom{A}}}^+}\,{7_{l}^-})$&&
 & $\;   -0.901749-i\,    0.997556\;$\\
$\;r_L^{[1/2]}({1_\qb^+}\,{5_g^+}\, {4_Q^+}\,{3_\Qb^-}\,{2_q^-}\,{6_{\lb_{\vphantom{A}}}^+}\,{7_{l}^-})$
&$\;0.00000\;$
 & $\;    -0.666667-i\,    0.000000$ & $\;    -2.527058-i\,    0.104842\;$\\
\hline
$\;A^{\tree}({1_\qb^+}\,{5_g^-}\,{4_Q^+}\,{3_\Qb^-}\,{2_q^-}\,{6_{\lb_{\vphantom{A}}}^+}\,{7_{l}^-})$&&
 & $\;   -0.298533+i\,    0.147741\;$\\
$\;r_L^{[1/2]}({1_\qb^+}\,{5_g^-}\, {4_Q^+}\,{3_\Qb^-}\,{2_q^-}\,{6_{\lb_{\vphantom{A}}}^+}\,{7_{l}^-})$
&$\;0.00000\;$
 & $\;    -0.666667+i\,    0.000000$ & $\;    -2.366982+i\,    0.089047\;$\\
\hline
$\;A^{\tree}({1_\qb^+}\,{4_Q^+}\,{5_g^+}\,{3_\Qb^-}\,{2_q^-}\,{6_{\lb_{\vphantom{A}}}^+}\,{7_{l}^-})$&&
 & $\;   0.288607+i\,    0.043034\;$\\
$\;r_L^{[1/2]}({1_\qb^+}\, {4_Q^+}\,{5_g^+}\,{3_\Qb^-}\,{2_q^-}\,{6_{\lb_{\vphantom{A}}}^+}\,{7_{l}^-})$
&$\;0.00000\;$
 & $\;    -0.66666+i\,    0.000000$ & $\;    -2.656572+i\,    0.000000\;$\\
\hline
$\;A^{\tree}({1_\qb^+}\,{4_Q^+}\,{5_g^-}\,{3_\Qb^-}\,{2_q^-}\,{6_{\lb_{\vphantom{A}}}^+}\,{7_{l}^-})$&&
 & $\;   0.148841-i\,    0.001185\;$\\
$\;r_L^{[1/2]}({1_\qb^+}\, {4_Q^+}\,{5_g^-}\,{3_\Qb^-}\,{2_q^-}\,{6_{\lb_{\vphantom{A}}}^+}\,{7_{l}^-})$
&$\;0.00000\;$
 & $\;    -0.666667+i\,    0.000000$ & $\;    -2.656572+i\,    0.000000\;$\\
\hline
$\;A^{\tree}({1_\qb^+}\,{4_Q^+}\,{3_\Qb^-}\,{5_g^+}\,{2_q^-}\,{6_{\lb_{\vphantom{A}}}^+}\,{7_{l}^-})$&&
 & $\;   -0.734834+i\,    0.360895\;$\\
$\;r_L^{[1/2]}({1_\qb^+}\, {4_Q^+}\,{3_\Qb^-}\,{5_g^+}\,{2_q^-}\,{6_{\lb_{\vphantom{A}}}^+}\,{7_{l}^-})$
&$\;0.00000\;$
 & $\;    -0.666667+i\,    0.000000$ & $\;    -2.887087-i\,    0.164880\;$\\
\hline
$\;A^{\tree}({1_\qb^+}\,{4_Q^+}\,{3_\Qb^-}\,{5_g^-}\,{2_q^-}\,{6_{\lb_{\vphantom{A}}}^+}\,{7_{l}^-})$&&
 & $\;   -0.421057-i\,    0.463392\;$\\
$\;r_L^{[1/2]}({1_\qb^+}\, {4_Q^+}\,{3_\Qb^-}\,{5_g^-}\,{2_q^-}\,{6_{\lb_{\vphantom{A}}}^+}\,{7_{l}^-})$
&$\;0.00000\;$
 & $\;    -0.666667+i\,    0.000000$ & $\;    -2.749917+i\,    0.137009\;$\\
\hline
$\;A^{\tree}({1_\qb^+}\,{4_Q^+}\,{3_\Qb^-}\,{2_q^-}\,{5_g^+}\,{6_{\lb_{\vphantom{A}}}^+}\,{7_{l}^-})$&&
 & $\;   1.347977+i\,    0.593626\;$\\
$\;r_L^{[1/2]}({1_\qb^+}\, {4_Q^+}\,{3_\Qb^-}\,{2_q^-}\,{5_g^+}\,{6_{\lb_{\vphantom{A}}}^+}\,{7_{l}^-})$
&$\;0.00000\;$
 & $\;    -0.666667+i\,    0.000000$ & $\;    -2.662151+i\,    0.000000\;$\\
\hline
$\;A^{\tree}({1_\qb^+}\,{4_Q^+}\,{3_\Qb^-}\,{2_q^-}\,{5_g^-}\,{6_{\lb_{\vphantom{A}}}^+}\,{7_{l}^-})$&&
 & $\;   0.570749+i\,    0.316836\;$\\
$\;r_L^{[1/2]}({1_\qb^+}\, {4_Q^+}\,{3_\Qb^-}\,{2_q^-}\,{5_g^-}\,{6_{\lb_{\vphantom{A}}}^+}\,{7_{l}^-})$
&$\;0.00000\;$
 & $\;    -0.666667+i\,    0.000000$ & $\;    -2.662151+i\,    0.000000\;$\\
\hline
\end{tabular} \end{center} \caption{
The primitive tree-level  four-fermion amplitude $A^{\tree}$ and
the ratio $r_L^{[1/2]}$ of the primitive one-loop amplitude  to primitive 
tree-level amplitude  for various helicities.}
\label{tab10}
\end{table}

\end{document}